\documentclass[fleqn,usenatbib]{mnras}

\usepackage{newtxtext,newtxmath}

\usepackage[T1]{fontenc}
\usepackage{ae,aecompl}
\usepackage{pdflscape} 

\usepackage{graphicx}	
\usepackage{amsmath}	
\usepackage{amssymb}	
\usepackage{adjustbox}
\usepackage{multirow}

\title[Formation History of the Milky Way disc]{The Formation History of the Milky Way disc with high-resolution cosmological simulations}

\author[Giammaria et al.]
{Marco Giammaria,$^{1,2}$\thanks{E-mail: marco.giammaria@inaf.it}
Alessandro Spagna,$^{1}$\thanks{E-mail: alessandro.spagna@inaf.it}
Mario G. Lattanzi,$^{1}$ 
Giuseppe Murante,$^{5}$
\newauthor{Paola Re Fiorentin,$^{1}$ Milena Valentini$^{3,4,5}$}
\\
$^{1}$INAF - Osservatorio Astrofisico di Torino, Via Osservatorio 20, I-10025, Pino Torinese, Turin, Italy \\
$^{2}$Department of Physics, University of Turin, Via P. Giuria, 1, I-10125 Turin, Italy\\
$^{3}$Universit{\"a}ts-Sternwarte M{\"u}nchen, Fakult{\"a}t f{\"u}r Physik, LMU Munich, Scheinerstr. 1, 81679 M{\"u}nchen, Germany\\
$^{4}$Excellence Cluster ORIGINS, Boltzmannstr. 2, D-85748 Garching, Germany\\
$^{5}$INAF - Osservatorio Astronomico di Trieste, via Tiepolo 11, I-34131 Trieste, Italy}

\date{Accepted XXX. Received YYY; in original form ZZZ}

\pubyear{2020}

\begin{document}
\label{firstpage}
\pagerange{\pageref{firstpage}--\pageref{lastpage}}
\maketitle

\begin{abstract}
    We analyse from an observational perspective the formation history and kinematics of a Milky Way-like galaxy from a high-resolution zoom-in cosmological simulation that we compare to those of our Galaxy as seen by Gaia DR2 to better understand the origin and evolution of the Galactic thin and thick discs. The cosmological simulation was carried out with the GADGET-3 TreePM+SPH code using the MUlti Phase Particle Integrator (MUPPI) model. We disentangle the complex overlapping of stellar generations that rises from the top-down and inside-out formation of the galactic disc. We investigate cosmological signatures in the phase-space of mono-age populations and highlight features stemming from past and recent dynamical perturbations. In the simulation, we identify a satellite with a stellar mass of $1.2 \cdot 10^9~\rm{M}_\odot$, i.e. stellar mass ratio $\Delta \sim 5.5$ per cent at the time, accreted at $z \sim 1.6$, which resembles the major merger Gaia-Sausage-Enceladus that produced the Galactic thick disc, i.e. $\Delta \sim 6$ per cent. We found at $z \sim 0.5$ -- 0.4 two merging satellites with a stellar mass of $8.8 \cdot 10^8~\rm{M}_\odot$ and $5.1 \cdot 10^8~\rm{M}_\odot$ that are associated to a strong starburst in the Star Formation History, which appears fairly similar to that recently found in the Solar Neighbourhood. Our findings highlight that detailed studies of coeval stellar populations kinematics, which are made available by current and future Gaia data releases and in synergy with simulations, are fundamental to unravel the formation and evolution of the Milky Way discs. 
\end{abstract}




\maketitle


\section{Introduction}
\label{sec:intro}
	
    Many of the chemo-kinematic properties of the Milky Way (MW) and nearby galaxies are related to events that occurred a long time ago. The aim of Local Cosmology is to provide a link between the local scenario at redshift $z=0$ and the distant Universe \citep[e.g.][]{fbh02}.

    In the era of precision astrometry, opened by the impressive improvement given by the Gaia Second Data Release \citep[DR2,][]{gaiamission}, the comparison between observations and high-resolution cosmological simulations of MW-like disc galaxies becomes mandatory to create a coherent laboratory for Local Cosmology studies. The standard cosmological $\Lambda$CDM model predicts that our Galaxy formed through the hierarchical merging of substructures, whose accretion history can be inferred from the chemo-kinematic signatures that we observe today in the stellar populations of the Galactic bulge, halo, thick and thin discs. With the advent of cosmological hydrodynamical zoom-in simulations of MW-like galaxies, we are able to analyse in detail the complex interplay of physical processes (e.g. dynamics of collisionless systems, gas flows, star formation, stellar nucleosynthesis, ...) that generated our Galaxy, which represents the Rosetta stone of galaxy evolution.

    Recently, several studies have confirmed that our Galaxy experienced $\sim 10$~Gyr ago a major merger with a massive dwarf galaxy, named Gaia-Sausage-Enceladus \citep[GSE, ][]{bel18, helmi18, dm19, vincenzo19, gal19}. According to these studies, the stellar inner halo is mainly made up of stellar debris from the accreted satellite and heated disc stars. However, the actual origin of the present thick disc is not clear and it is a still matter of debate whether it derives mainly from the kinematic heating of the proto-disc, or from a starburst triggered by merging events and fed by the infall of a gas-rich satellite \citep{brook04, brook12}. An extensive review of the early accretion history of the MW is described by \citet{helmi20}, while detailed comparisons with the chemo-dynamical properties of cosmological simulations are presented by \citet{minchev13}, \citet{s13}, \citet{buck19} and \citet{grand20}.
    
    Other authors have also produced and analysed high resolution simulations in order to reconstruct the formation history of our MW; some of these studies are listed in Table~\ref{tab:cosmological_simulations}.

    Here, we provide a detailed analysis of Aquila-C-4 (AqC4), a hydrodynamical cosmological simulation of a MW-mass galaxy based on the MUPPI algorithm \citep{murante10, murante15}. The AqC series of simulations has been presented by \citet{spr08} and used, among other papers, in the Aquila comparison project \citep{sc12}. The validity of the MUPPI algorithm has been confirmed by several studies \citep[e.g.][]{monaco12, goz15, MV17}. It consists of an unconstrained simulation, i.e. it is not designed to exactly mimic the dynamic history of our Galaxy as in simulations aimed to reproduce the Local Group environment \citep[see for example][]{sawala16, carlesi16}. Thus, according to the precepts of Local Cosmology, AqC4 results as a typical cosmological product with mass and phase-space properties similar to the MW, and with a merging history comparable to a disc-like galaxy.
    
    In this paper, we aim to compare AqC4, namely a theoretical `error-free' catalogue predicted for a Milky Way-like galaxy, to our Galaxy as seen by Gaia DR2 \citep{gaiamission}. Therefore, the analysis presented in this work is carried out as if the simulation were a real stellar survey of our Galaxy and we focus on the phase-space properties of stellar particles contained in a simulated region representative of the Solar Neighbourhood. We characterize the spatial and kinematical parameters of the stellar disc and investigate the Star Formation History (SFH) of AqC4 for redshift $0 < z \lesssim 2$, in order to identify signatures of past and recent merging events that can be used to disentangle the accretion history of our Galaxy.
    
    In Sect.~\ref{sec:simulation} we describe the simulation and summarize its main parameters. In Sect.~\ref{sec:spadist} we characterize the spatial distribution of particles, focusing on the radial and vertical structures of the stellar disc, i.e. determining its scale length and scale height. We investigate the mono-age populations in order to study the flaring of the stellar disc. In Sect.~\ref{sec:kinprop} we analyse the kinematic properties of AqC4. We study the global rotation curve of AqC4 and we make a direct comparison with recent observational data of the MW as seen by Gaia. Then, we focus on the substructures of the stellar disc in the region defined as the Simulated Solar Ring (hereafter SSR). In Sect.~\ref{sec:galevol} we link our previous findings to the formation and evolution of AqC4; specifically, we extensively compare the Star Formation History of the simulation with recent estimates for the MW, and then we extend our investigation to the accretion history. Finally, in Sect.~\ref{sec:concl} we summarize our findings and discuss our results.
	
    \begin{table}
        \centering
        \caption{Cosmological hydrodynamical simulations of MW-like disc galaxies. Main properties of recent high resolution zoom-in simulations.}
        \label{tab:cosmological_simulations}
        \begin{adjustbox}{width=\columnwidth,center}
        \begin{tabular}{lllr}
            \hline
            Project & Mass particle & Softening & Reference \\
                    & [M$_\odot$] & [pc] & \\
            \hline
            Eris & M$_{\rm DM}\sim 1 \cdot 10^5$ & $\epsilon_*\sim 120$ & \citet{guedes11} \\
                    & M$_{\rm gas}\sim 2\cdot 10^4$ & $\epsilon_{\rm gas}\sim 120$ & \\
            \hline
	        Auriga & M$_{\rm DM}\sim 3\cdot 10^5$ & $\epsilon_*\sim 369$ & \citet{g17} \\
                    & M$_{\rm gas}\sim 5\cdot 10^4$ &  & \\
            \hline
            GIZMO & M$_{\rm DM}\sim 3\cdot 10^5$ & $\epsilon_*\sim 50$  & \citet{ma17} \\
                    & M$_{\rm gas}\sim 6\cdot 10^4$ & $\epsilon_{\rm gas}\sim 14$ & \\
            \hline
            Illustris & M$_{\rm DM}\sim 7\cdot 10^6$ & $\epsilon_* \sim 740$ & \citet{nel18} \\
            TNG100 & M$_{\rm gas}\sim 1\cdot 10^6$ & $\epsilon_{\rm gas} \gtrsim 185$ & \\
            \hline
            EAGLE & M$_{\rm DM}\sim 1\cdot 10^6$ &  & \citet{mac19} \\
                    & M$_{\rm gas}\sim 2\cdot 10^5$ & $\epsilon_{\rm gas} \lesssim 350$ & \\
            \hline
            NIHAO-UHD & M$_{\rm DM}\sim 1\cdot 10^5$ & $\epsilon_*\sim 273$  & \citet{buck20} \\
            (g7.08e11) & M$_{\rm gas}\sim 2\cdot 10^4$ & $\epsilon_{\rm gas}\sim 273$ & \\
            \hline
            AqC4 & M$_{\rm DM}\sim 4\cdot 10^5$ & $\epsilon_*\sim 223$  & This work \\
             & M$_{\rm gas}\sim 7\cdot 10^4$ & $\epsilon_{\rm gas}\sim 223$ & \\
            \hline
        \end{tabular}
        \end{adjustbox}
    \end{table}

	
\section{Simulation}
\label{sec:simulation}
	
    In this Section, we introduce the simulation. The cosmological hydrodynamical simulation that we performed has been carried out with the TreePM+SPH (smoothed particle hydrodynamics) GADGET3 code, a non-public evolution of the GADGET2 code \citep{Springel2005}. As initial conditions (ICs), we adopt the zoomed-in ICs AqC4 introduced by \citet{spr08}. In our realization, they describe an isolated DM halo at redshift $z = 0$, with a virial mass\footnote{The virial quantities are those calculated in a sphere that is centred on the minimum of the gravitational potential of the halo and that encloses an over-density of 200 times the critical density at present time.} of $M_{\rm vir} \sim 1.627 \cdot 10^{12}$~M$_\odot$ within a spherical volume of $R_{\rm vir} = 237.13$~kpc. Current estimates of MW virial mass are consistent with the range of values 0.5 -- $2 \cdot 10^{12}$~M$_\odot$ \citep{bg16}. In particular, studies based on Gaia DR2 give $M_{\rm vir} \sim 1.3 \cdot 10^{12}$~M$_\odot$ when considering the dynamics of globular clusters \citep{ph19, wat19}, and $M_{\rm vir} \sim 1 \cdot 10^{12}$~M$_\odot$ using MW rotation curve \citep{e19, crosta20}. In what follows, we consider the galactic radius, $R_{\rm gal}$, as $1/10$ of the virial one\footnote{We count a total number of DM, gas and stellar particles respectively of $5\,518\,587$, $1\,348\,120$, and $6\,919\,646$ at redshift $z=0$ within the virial volume.}.

    We assume a $\Lambda$CDM cosmology with $\Omega_{\rm m}=0.25$, $\Omega_{\rm \Lambda}=0.75$, $\Omega_{\rm baryon}=0.04$, $\sigma_8=0.9$, $n_s=1$, and $H_{\rm 0}=100\,h$~km~s$^{-1}$~Mpc$^{-1}=73$~km~s$^{-1}$~Mpc$^{-1}$ where $h$ represents the reduced Hubble constant. The zoomed-in region that we simulate has been extracted from a cosmological volume of $100 \, (h^{-1}$~Mpc$)^{3}$ of the DM-only, parent simulation.
    
    The Plummer-equivalent softening length for the computation of the gravitational force is $\epsilon_{\rm Pl} = 163~h^{-1}$~pc. DM particles have a mass of $2.7 \cdot 10^5~h^{-1}$~M$_\odot$, and the initial mass of gas particles is $5.1 \cdot 10^4~h^{-1}$~M$_\odot$.

    As anticipated in Sect.~\ref{sec:intro}, we adopt a sub-resolution model called MUPPI (MUlti Phase Particle Integrator). Here we outline its most relevant features, while we refer the reader to \citet{murante10, murante15} for a more comprehensive description. In particular, all the parameters of the model which are not explicitly mentioned here are set to the same values as in \citet{murante15}. The MUPPI sub-resolution model describes a multi-phase interstellar medium (ISM). Among the most important processes implemented, it features star formation and stellar feedback, metal cooling and chemical evolution. It accounts for hot and cold ($T=300$~K) gas phases in pressure equilibrium by means of multiphase particles, which are the building blocks of the model. A gas particle enters a multi-phase stage if its density increases above a density threshold ($n=0.01$~cm$^{-3}$) and its temperature drops below a temperature threshold ($T=10^5$~K). 

    A set of ordinary differential equations describes mass and energy flows among different components within each multiphase particle: for instance, radiative cooling moves mass from the hot into the cold phase, while a tiny fraction of the cold gas evaporates due to the destruction of molecular clouds. We rely on the phenomenological prescription by \citet{Blitz2006} to estimate the fraction of cold gas which is in the molecular phase and which fuels star formation. Star formation is then modelled according to the stochastic algorithm introduced by \citet{SpringelHernquist2003}: as a consequence, a multiphase gas particle can generate (up to four generations of) star particles. 

    The MUPPI model features stellar feedback both in thermal and kinetic forms \citep[as described in][]{murante15}. Besides stellar feedback in energy, star formation and evolution also result in a chemical feedback, and galactic outflows generated by supernova (SN) explosions promote metal spread and circulation within the galaxy \citep{MV17, Valentini2018}.

    Our model accounts for stellar evolution and chemical enrichment following \citet{tornatore2007}, where a thorough description can be found. In summary, each star particle is considered to be a simple stellar population. Assuming an initial mass function \citep{Kroupa1993}, as well as predictions for stellar lifetimes \citep{PadovaniMatteucci1993} and stellar yields \citep[see][for details]{murante15}, we evaluate the number of stars aging and eventually exploding as SNe, and the amount of metals injected in the ISM. Heavy elements released by star particles are distributed to neighbouring gas particles. The chemical evolution of $9$ metals (C, Ca, O, N, Ne, Mg, S, Si, Fe) synthesized by different sources (namely asymptotic giant branch stars, SNe~Ia and SNe~II) is individually followed. The model also features metallicity-dependent radiative cooling \citep[following][]{Wiersma2009}, and includes the effect of an ionizing cosmic background \citep{HaardtMadau2001}.
		
	
\section{Spatial distribution of the disc}
\label{sec:spadist}
	
\begin{figure*}
		\centering
		\includegraphics[width=\linewidth]{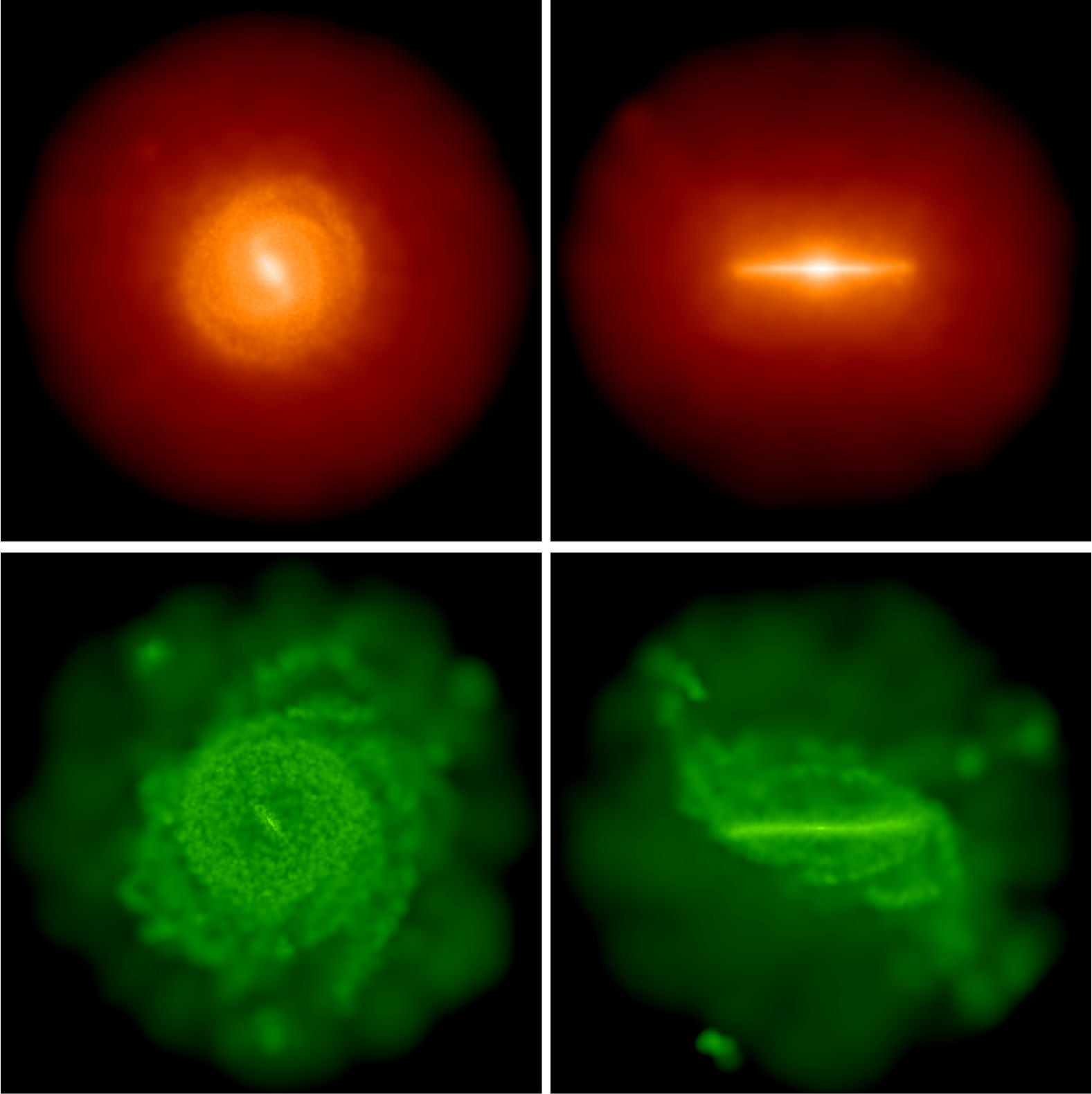}
		\caption{Stellar (upper panels) and gas (lower panels) projected density for the AqC4 simulation (face-on and edge-on view on left and right panels, respectively). The $Z$-axis of the coordinate system is aligned with the angular momentum vector of multi-phase gas and stars enclosed within 8~kpc from the position of the minimum of the gravitational potential. The total box size is 50~kpc.}
		\label{fig:AqC4_dens}
	\end{figure*}

	Fig.~\ref{fig:AqC4_dens} shows the face-on and edge-on projected density of stellar (upper panels) and gas (lower panels) particles. The reference system used is centred in the minimum of the gravitational potential and rotated to be aligned with the angular momentum vector of multi-phase gas and star particles within 8~kpc from the centre, while the $XY$ plane is orthogonal to it. This reference system has been adopted for the entire analysis presented in this paper.
	
	The presence of a disc structure is clear both in the stellar and gas distribution. This disc dominates the central part of the galactic volume and is extended until $R \sim 10$~kpc. In the innermost central region, a non-axisymmetric bar structure is visible in both components. The stellar disc is quite symmetric around the $XY$ plane which can be defined itself as the galactic plane. Finally, a spiral pattern is visible in the outer part of the disc region, while the gas exhibits a warped shape at its edge.
	
\begin{figure}
		\centering
		\includegraphics[width=\linewidth]{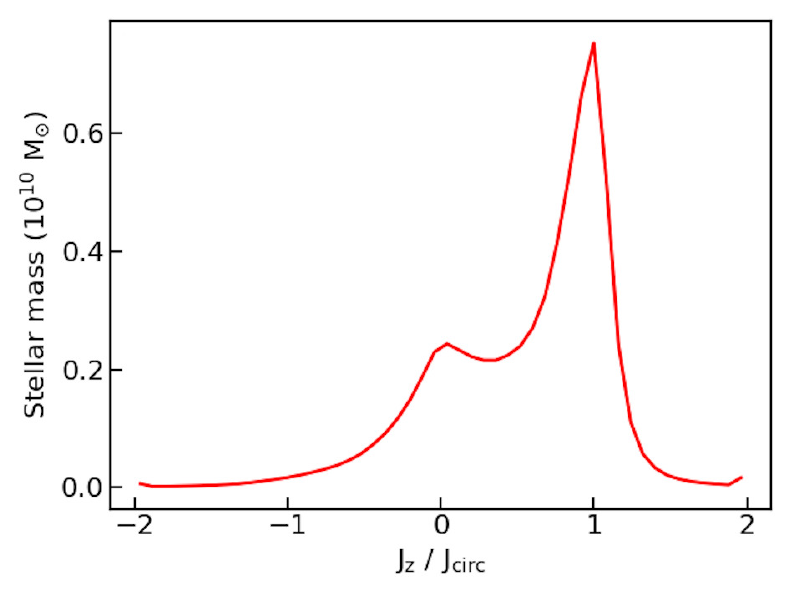}
		\caption{Distribution of the stellar mass of AqC4 as a function of the orbit circularity for all stellar particles within $R_{\rm gal}$ at redshift $z = 0$.}
		\label{fig:circularity}
	\end{figure}

	We characterize the main components of AqC4 by means of the stellar mass distribution as a function of the orbit circularity of all star particles within $R_{\rm gal}$ for our simulation at redshift $z = 0$. The circularity of an orbit is defined as $\epsilon = J_{\rm Z}/J_{\rm circ}$, where $J_{\rm Z}$ is the specific angular momentum in the direction perpendicular to the disc, and $J_{\rm circ}$ is the specific angular momentum of a reference circular orbit. The results are shown in Fig.~\ref{fig:circularity}.
	
	The prominent peak at $\epsilon \sim 1$ demonstrates that AqC4 is characterized by a disc structure, with a bulge component corresponding to the smaller peak at $\epsilon \sim 0$. We compute the ratio of bulge-over-total stellar mass $B/T$ by doubling the mass of the counter-rotating stars within $R_{\rm gal}$, under the hypothesis that the bulge is supported by velocity dispersion and thus has an equal amount of co- and counter-rotating stars. We highlight that this is a `dynamical' value of the bulge-to-total stellar mass ratio as defined by \citet{scannapieco10}. The resulting ratio is $B/T = 0.34$, fairly comparable with the upper limit estimated for the MW, $B/T_{\rm MW} \simeq 0.15$ -- 0.33 \citep[see e.g.][and references therein]{bg16, bell17}.
	
\subsection{Radial scale length of the stellar disc}
\label{sec:hr}
	
	It is common to describe the stellar disc of the MW with a double-component exponential decay profile with a radial scale length of $h_{\rm Rt} = 2.6 \pm 0.5$~kpc for the thin disc and $h_{\rm RT} = 2.0 \pm 0.2$~kpc for the thick disc \citep{bg16}. 
	
\begin{figure}
		\centering
		\includegraphics[width=\linewidth]{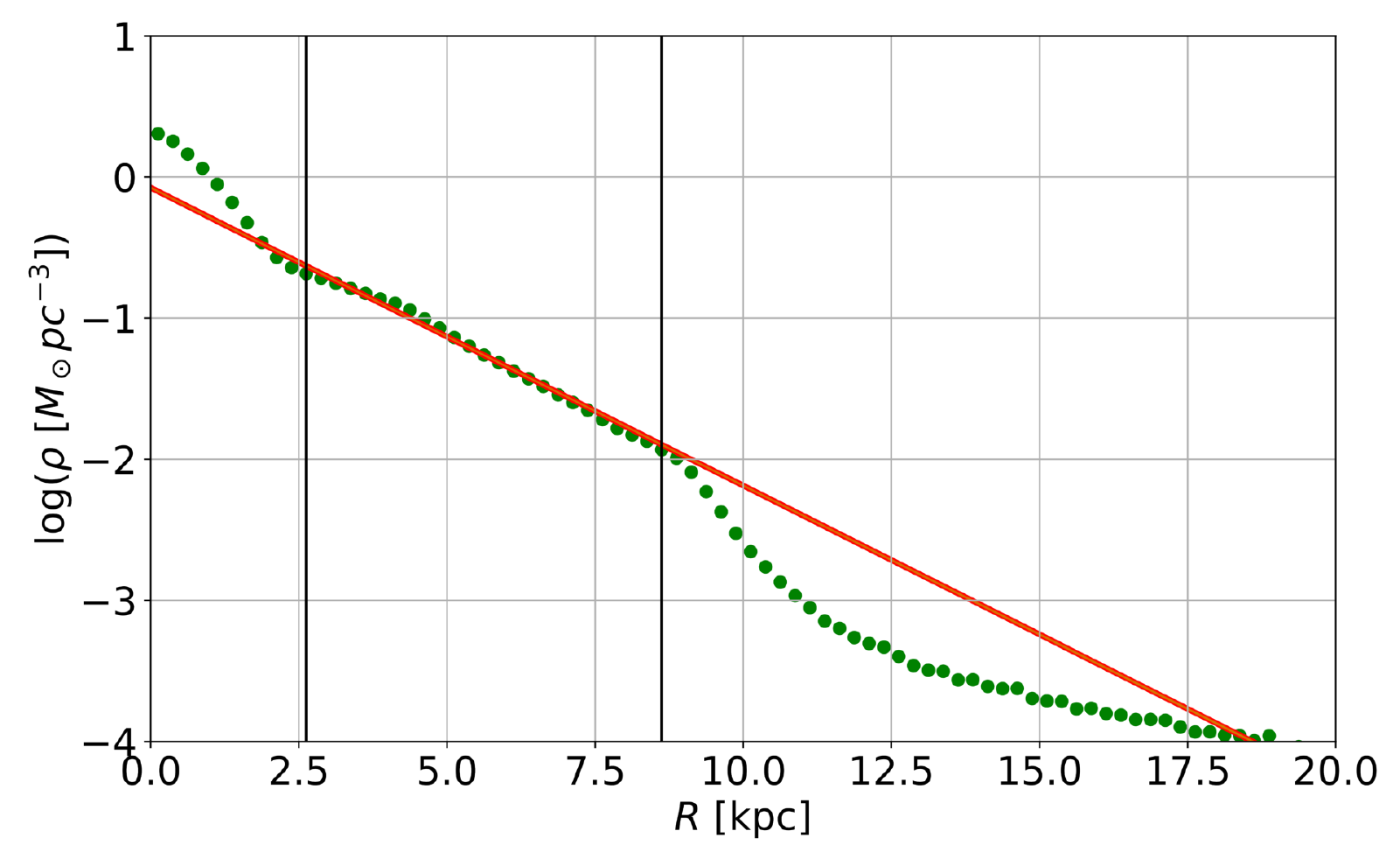}
		\caption{Radial density distribution (green dots) of stellar particles within $|Z| \leq 1$~kpc from the galactic plane computed in radial bins of $\Delta R = 0.25$~kpc. Error bars are smaller than the size of data point. Red line represents the best fit obtained from the posteriors of the MCMC analysis. The CI is thinner than the best-fitting red line. Vertical black lines show the radial interval adopted for the fit.}
		\label{fig:hr}
	\end{figure}

	In order to determine the radial extension of the stellar disc of AqC4, we first select all the star particles that lie within a height on the galactic plane $|Z| \leq 1$~kpc and compute the volume mass density (in $\rm{M}_\odot \rm{pc}^{-3}$) for cylindrical radial bins of $\Delta R = 0.25$~kpc as shown in Fig.~\ref{fig:hr}. Then, to infer the scale length, we limit the fit in the disc-dominated region between $2.5 \leq R [\rm{kpc}] \leq 9$ (solid black lines).
	
	Note that for this and the following analysis, we implement an MCMC Bayesian algorithm using the Python package PyMC3 \citep{pymc3} to take into account the mass of stellar particles, the weighted errors due to the Poissonian statistics within each bin and to check if the uncertainties induced by the bin size could be influenced by the softening resolution limit of the simulation, i.e. $\epsilon=0.163$~kpc.
	
	We analyse the stellar particles close to the galactic plane, i.e. $|Z| \leq 1$~kpc, and estimate\footnote{We set the Credible Interval (CI) as the $\pm 1\sigma$ equivalent range between the 16$^{\rm{th}}$ and 84$^{\rm{th}}$ percentile of the posterior distribution, i.e. the percentage equivalents of the $1\sigma$ range in a Gaussian distribution. From the posteriors, it results very narrow and thinner than the best-fitting line.} a radial scale length of $h_{\rm R} = 2.058 \pm 0.002$~kpc that is smaller by a factor of $\sim 21$ per cent with respect to the MW thin disc. Thus, in the following we adopt a proportionally smaller Simulated Solar Ring (SSR) of $6 \leq R [\rm{kpc}] \leq 7$ with respect to the distance between the Sun and the Galactic Centre, $R_\odot = 8.122 \pm 0.033$~kpc, as estimated by the \citet{gravity}.
	
\subsection{Vertical distribution of the stellar disc in the SSR}
\label{sec:hzSSR}
	
	The vertical mass distribution of the MW disc is modelled as a double-component exponential function.
	 
	As reported by \citet{bg16}, the thin disc scale height at solar distance from the Galactic Centre is $h_{\rm Zt} = 300 \pm 50$~pc, while for the thick disc is $h_{\rm ZT} = 900 \pm 180$~pc. Moreover, several studies have tried to determine the relative density normalization $f = \rho_{\rm T}/\rho_{\rm t}$ of the thick disc compared to the thin disc with estimates ranging from 6 to 12 per cent \citep[e.g.][]{j08, jj10, b15}. 
	
	In order to reduce the contamination from halo stars, we select all the $320\,354$ stellar particles in the SSR within $|Z| \leq 3$~kpc and fit simultaneously a double-component disc described as follows:
	
	\begin{equation}
	\rho(Z) = A \Bigg(e^{-\frac{|Z|}{h_1}} + f\cdot e^{-\frac{|Z|}{h_2}}\Bigg),
	\label{eq:rhoZ}
	\end{equation}
	where $A$ is the total density normalization, $f$ is the relative density normalization, $h_i$, for $i = 1,2$, is the scale height of the two components of the disc, and $Z$ is the vertical coordinate. 
	
\begin{figure}
		\centering
		\includegraphics[width=\linewidth]{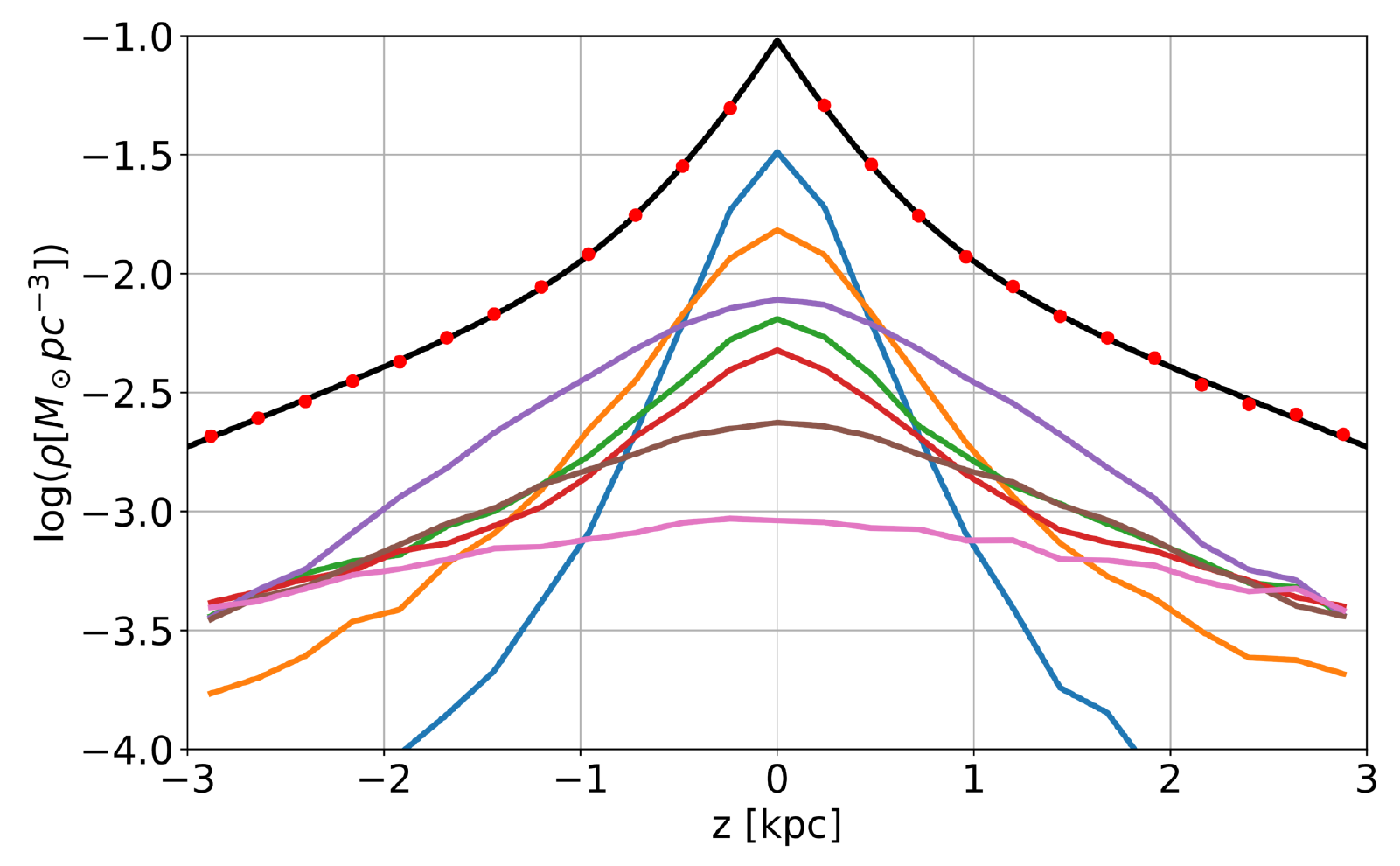}	
		\caption{Vertical density distribution (red dots) of stellar particles in the SSR with $|Z| \leq 3$~kpc computed in vertical bins of $\Delta Z = 0.25$~kpc. Black line represents the best fit obtained from the posteriors of MCMC analysis. The CI is thinner than the best-fitting black line. Colour lines represent mono-age stellar populations divided in bins of 2~Gyr as listed in Table~\ref{tab:mono-age_hz}.}
		\label{fig:hz}
	\end{figure}

	We estimate the scale heights, $h_1 = 0.305 \pm 0.005$~kpc that is in good agreement with the MW thin disc, and $h_2 = 1.33 \pm 0.02$~kpc that results $\sim 50$ per cent larger than what it is estimated for the MW thick disc (see Fig.~\ref{fig:hz}). We also found a relative density parameter of $f=22.6 \pm 0.7$ per cent that is 2-3 times larger with respect to the MW \citep{bg16}. These differences derive from the diverse accretion history of the MW and AqC4, as described in Sect.~\ref{sec:galevol}. 
		
	A better understanding of the stellar disc distribution is given by the stellar sub-samples formed in 2~Gyr time bins (see Table~\ref{tab:mono-age_hz}) shown as coloured lines in Fig.~\ref{fig:hz}. We can distinguish four main components: (i) the {\it young disc} population with stellar ages up to 4~Gyr (blue and orange lines) that shows an exponential decreasing from the galactic plane similar to the {\it thin disc} described in Sect.~\ref{sec:hzSSR}; (ii) an intermediate {\it old disc} population (green and red lines, i.e. 4-8~Gyr old), whose vertical distribution clearly shows the superposition of a thin and a thick component; (iii) a prominent {\it thick disc} population in the age interval 8-10~Gyr (purple line) corresponding to the bulk of the `geometric' thick disc; (iv) the spheroidal {\it oldest population} with age greater than 10~Gyr (brown and pink lines) that shows widespread distribution, and includes a large fraction of halo star particles.
	
	We point out that 40 per cent of the sample is made up of \textit{young disc} stars with ages $\leq 4$~Gyr that dominate the stellar population in the SSR. The sum of the previous two bins (i.e. $4 \leq \rm{Age} [\rm{Gyr}] \leq 8$) attains only 22 per cent of the sample, while the same fraction is found in the antecedent bin, $8 \leq \rm{Age} [\rm{Gyr}] \leq 10$, corresponding to the \textit{thick disc}.
	
	These results reveal a SFH affected by significant changes over time, which are discussed in more detail in Sect.~\ref{sec:SFH}.
	
	\begin{table}
		\caption{Age and relative weights of the mono-age populations with respect to the total stellar particles ($N_0$) in the SSR ($6 \leq R [\rm{kpc}] \leq 7$) within $|Z| \leq 3$~kpc.}
		\label{tab:mono-age_hz}
		\begin{center}
			\begin{tabular}{lcc}
				\hline
				Colour & Age [Gyr] & $N/N_0$ \\
				\hline
				blue & 0-2 & 0.22 \\
				orange & 2-4 & 0.18 \\
				green & 4-6 & 0.12 \\
				red & 6-8 & 0.10 \\
				purple & 8-10 & 0.22 \\
				brown & 10-12 & 0.10 \\
				pink & 12-14 & 0.07 \\
				\hline
				black & 0-14 & $N_0$=$320\,354$ \\
				\hline
			\end{tabular}
		\end{center}
	\end{table}

	
\subsection{The radial disc flaring}
\label{sec:flaring}
	
	Another important spatial feature which stems from the inside-out is the {\it disc flaring} \citep[see e.g.][ and references therein]{m17}. Here, we investigate this property by studying the vertical distribution of AqC4 stellar particles as a function of the cylindrical radial distance for both the whole sample and for the mono-age populations listed in Table~\ref{tab:mono-age_hz}.
	
\begin{figure}
		\centering
		\includegraphics[width=\linewidth]{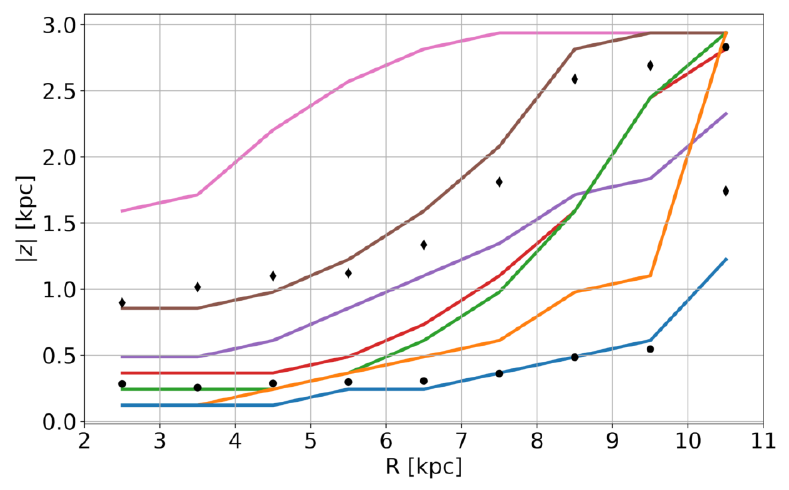}	
		\caption{Variation of stellar height, $|Z|$, with $R$. Flaring is present for all mono-age populations (colour code as in Table~\ref{tab:mono-age_hz}). Solid lines show when the density decays by a factor of e$^{-1}$. The $h_1$ and $h_2$ scale heights for each radial bin integrated over all stellar ages (see Table~\ref{tab:Rbin_hz}) are represented by circles and diamonds, respectively.}
		\label{fig:flaring}
	\end{figure}

	As shown in Fig.~\ref{fig:flaring}, we divided the star particles in rings of $\Delta R = 1$~kpc and $|Z| \leq 3$~kpc. In each ring we fit the vertical density distribution of the whole stellar sample with the double exponential model of Eq.~(\ref{eq:rhoZ}) and determine the variation of $h_1$ and $h_2$ parameters along the disc. Then, we estimate the thickness of each mono-age population as the height, $h_Z$,  where the density distribution decreases by a factor e$^{-1}$ with respect to the galactic plane, $Z = 0$~kpc.
	
	The results of Table~\ref{tab:Rbin_hz} report that flaring is present for all mono-age populations and occurs at smaller radii for older populations due to the inside-out formation of the disc. In particular, the youngest population ($\rm{Age} \leq 2$~Gyr) lies close to the galactic plane ($h_Z \lesssim 0.5$~kpc) along the whole disc up to $R \simeq 9$~kpc (cfr. Fig.~\ref{fig:hr}), while the intermediate {\it old-thin disc} population ($4 \leq \rm{Age} [\rm{Gyr}] \leq 8$) shows a significant flaring already at $R \simeq 6$~kpc with a steep increase towards larger radii. We recall that in the SSR the stellar particles appear described by double-component exponential vertical profiles (Fig.~\ref{fig:hz}): the presence of coeval stars evidences that heterogeneous populations were possibly formed by different progenitors.
	
	The prominent {\it thick disc} population (i.e. $8 \leq \rm{Age} [\rm{Gyr}] \leq 10$) shows a thickness $h_Z \simeq 1.0-1.2$~kpc in the SSR at $R=6-7$~kpc that is pretty close to the scale height $h_2 \simeq 1.33$~kpc estimated in Sect.~\ref{sec:hzSSR}. A radial flaring is already apparent at $R \simeq 4$~kpc, but it shows a milder increase at larger radii than the younger populations discussed above. These stellar particles were born in the same period of the starburst at $z \simeq 1.5$ (see Sect.~\ref{sec:SFH}) and formed the primordial disc. In Sect.~\ref{sec:merger} we will see that this component may represent the signature of one of the most important mergers in the accretion history of AqC4.
	
	In summary, our analysis of the spatial distribution of AqC4 confirms that a two-disc decomposition is a `simplified' mathematical description of a more complex physical scenario which implies an overlapping of different mono-age profiles for several stellar generations. These results support the global top-down, inside-out disc formation model and are consistent with previous observations of the MW \citep[e.g.][where in the latter the authors used mono-abundance populations in the MW instead of mono-age ones]{mmb11, gd14, b16} and simulations \citep{s13, g17, MV19}.
	
	\begin{table}
		\caption{The $h_1$ and $h_2$ scale heights, and relative density normalization $f$ with the $1\sigma$ CI integrated over all the $N_*$ stellar particles within each radial bin.}
		\label{tab:Rbin_hz}
		\begin{center}
			\begin{tabular}{ccccc}
				\hline
				$R$ [kpc] & $h_1$ [kpc] & $h_2$ [kpc] & $f$ [\%] & $N_*$\\
				\hline
				\\[-0.85em]
				2-3 & $0.306^{+0.004}_{-0.003}$ & $0.897^{+0.005}_{-0.005}$ & $35.2^{+0.9}_{-0.9}$ & $767\,311$\\  
				\\[-0.5em]
				3-4 & $0.256^{+0.002}_{-0.002}$ & $1.016^{+0.007}_{-0.007}$ & $13.2^{+0.3}_{-0.3}$ & $687\,832$\\
				\\[-0.5em]
				4-5 & $0.288^{+0.002}_{-0.002}$ & $1.09^{+0.02}_{-0.02}$ & $12.2^{+0.4}_{-0.4}$ & $592\,207$\\
				\\[-0.5em]
				5-6 & $0.297^{+0.003}_{-0.003}$ & $1.12^{+0.02}_{-0.02}$ & $21.1^{+0.7}_{-0.5}$ & $436\,270$\\
				\\[-0.5em]
				6-7 & $0.305^{+0.005}_{-0.005}$ & $1.34^{+0.02}_{-0.02}$ & $22.6^{+0.7}_{-0.7}$ & $320\,354$\\
				\\[-0.5em]
				7-8 & $0.362^{+0.007}_{-0.006}$ & $1.81^{+0.05}_{-0.05}$ & $22.0^{+0.9}_{-0.9}$ & $234\,454$\\
				\\[-0.5em]
				8-9 & $0.485^{+0.007}_{-0.007}$ & $2.6^{+0.1}_{-0.1}$ & $19.9^{+1.1}_{-1.1}$ & $171\,674$ \\
				\\[-0.5em]
				9-10 & $0.55^{+0.02}_{-0.02}$ & $2.7^{+0.1}_{-0.1}$ & $46^{+2}_{-2}$ & $96\,159$\\
				\\[-0.5em]
				10-11 & $1.7^{+0.2}_{-0.2}$ & $2.8^{+0.7}_{-0.9}$ & $89^{+4}_{-4}$ & $46\,019$\\
				\\[-0.85em]
				\hline
			\end{tabular}
		\end{center}
	\end{table}
	
	
\section{Kinematic properties}
\label{sec:kinprop}
	
	Now, we extend the analysis of the galactic structure done in Sect.~\ref{sec:spadist} to the kinematics of the stellar populations `observed' at redshift $z=0$ in AqC4. 
	
	
    \subsection{Rotation Curve}
    \label{sec:RC}
	
    The galactic Rotation Curve (RC) constitutes one of the key features that characterize disc galaxies like the MW. The significant discrepancy existing between the empirical curves observed \citep[e.g.][]{sr01, e19, crosta20}, and theoretical circular velocity profiles derived form the baryonic mass distribution (considering test particles that would move in an axisymmetric gravitational potential $\Phi$ according to Newtonian theory) represents one of the main line of evidence of the presence of dark matter (DM) in haloes \citep[e.g.][]{i15, mc17, ds19}.
	
\begin{figure}
		\centering
		\includegraphics[width=\linewidth]{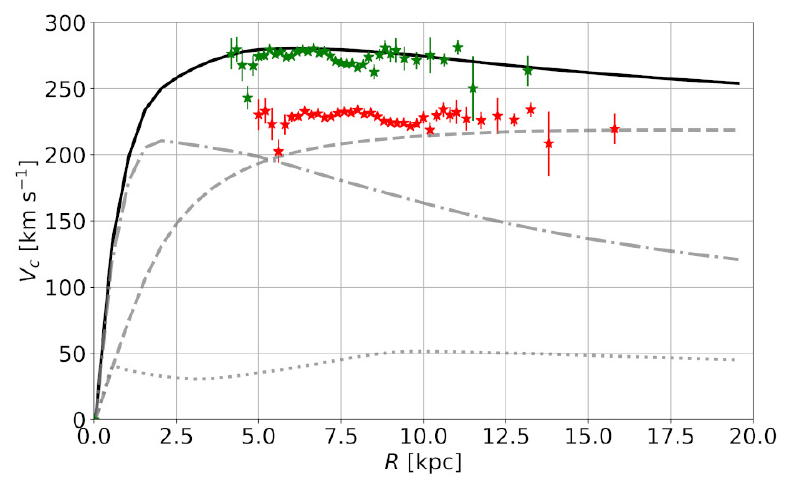}
		\caption{Rotation curves for AqC4. Solid line shows the total curve, while dashed, dot-dashed and dotted lines show the contribution of the DM, stellar and gas component, respectively. Red star symbols with corresponding uncertainties represent observational data for the MW from \citet{crosta20}. Green star symbols are the same Gaia DR2 data scaled as described in Sect.~\ref{sec:RC}.}
	 	\label{fig:RC}
	\end{figure}

	Fig.~\ref{fig:RC} shows the circular velocity $V_c(R)=\sqrt{R \partial\Phi / \partial R}=\sqrt{GM(<R)/R}$ for the total mass (black solid line), and for the individual component of DM (gray dashed), stars (gray dot-dashed), gas (dotted). As expected, after a steep linear increase for $R \leq 3$~kpc, the RC has an almost flat profile in the disc-dominated region for $3 \leq R [\rm{kpc}] \leq 11$, reaches its peak $V_c = 280.6 $~km~$\rm{s}^{-1}$ at $R = 6.25$~kpc, and finally has a smooth decrease for $R \geq 11$~kpc in the halo dominated region. In the central region, for $R \lesssim 5$~kpc, the RC is dominated by the mass of stellar particles, while for $R \gtrsim 5$~kpc the main contribution is due to DM. This is similar to what happens in the MW \citep[e.g. see Fig. 1 in][]{crosta20} taking into account the relative size of the MW disc which is more extended than the one of AqC4.
	
	In the SSR, the mean circular velocity is $280.1$~km~$\rm{s}^{-1}$, which is $\sim 20$ per cent faster than the value $V_\phi \simeq 234$~km~$\rm{s}^{-1}$ measured at the Sun position $R_\odot=8.122$~kpc by \citet{crosta20}. However, if we scale the size (with the radial scale length estimated in Sect.~\ref{sec:hr}) and the velocity, keeping \textit{constant} the angular momentum $L_Z$ at the Sun position, we can reasonably overplot the RC of the MW (green star symbols in Fig.~\ref{fig:RC}) on that of AqC4.
	
	Thus, AqC4 appears to be a disc galaxy fairly similar to the MW with a $\sim 20$ per cent faster rotating disc and a shorter \textit{pseudo}-solar position location. The re-scaled Gaia data that describe the Galactic disc kinematics results in good agreement with the rotation curve of AqC4.
		 
	\subsection{Multi-component disc in the SSR}
	\label{sec:TNM}
	
	In Sect.~\ref{sec:hzSSR}, we have modeled the stellar vertical distribution in the SSR with the superposition of a thin disc ($h_1 \simeq 0.305$~kpc) and thick disc ($h_2 \simeq 1.33$~kpc) that includes a residual contamination of halo star particles.
	
	Here, we focus on the rotation velocity distribution of the stellar particles within the SSR in order to identify and characterize the different components. This methodology is usually applied to real stellar surveys in order to select stellar samples belonging to the Galactic populations, e.g. the thin and thick disc, and inner halo \citep{bond10, spagna10}. Only a few authors have applied such kinematic decomposition to cosmological simulations \citep[see][and references therein]{abadi03a, abadi03b, obreja18, obreja19}, as the high-resolution level required has been only recently achieved.
	
\begin{figure*}
	    \centering
		\includegraphics[width=\linewidth]{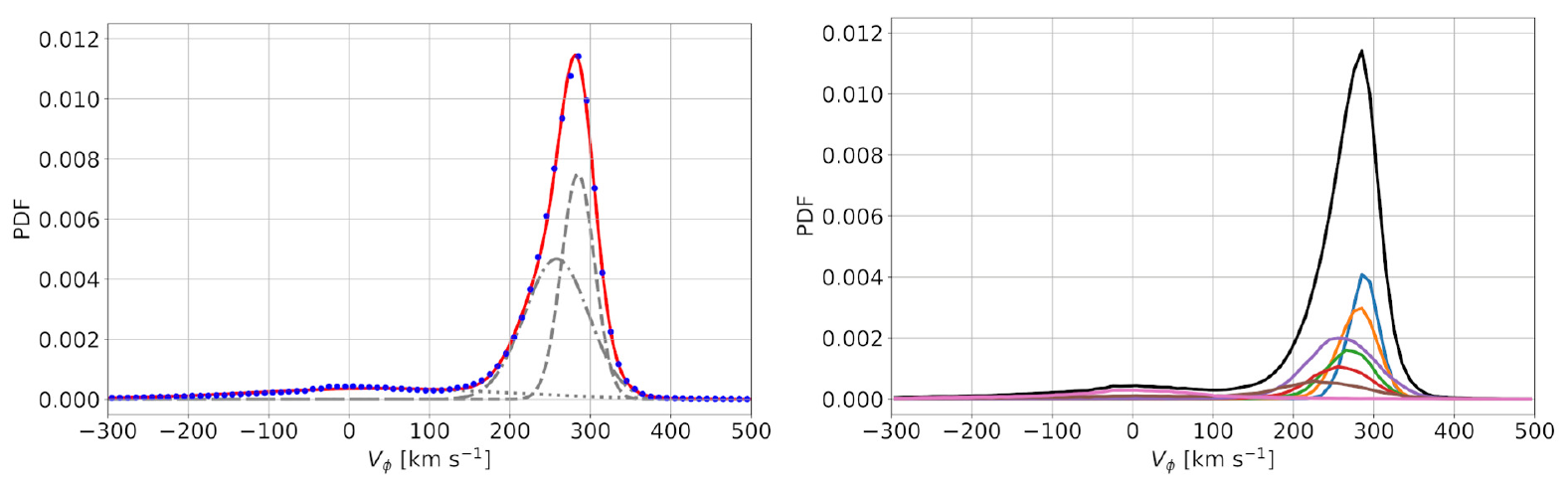}
		\caption{\emph{Left panel}: Kinematic decomposition of $V_{\rm \phi}$ PDF for stellar particles in the SSR (i.e. $6 \leq R \rm{[kpc]} \leq 7$) with $|Z|\leq 4$~kpc. The observed values in velocity bins of $\Delta V = 10$~km~$\rm{s}^{-1}$ are represented by blue dots. \textit{Young disc}, \textit{old disc/thick disc}, and halo structure are shown with dashed, dot-dashed and dotted lines, respectively. The reconstructed profile of the PDF using the posteriors estimates of the TNM model is shown by the red line. \emph{Right panel}: $V_{\rm \phi}$ distribution for mono-age stellar particles. Colour code as in Table~\ref{tab:mono-age_hz}.}
		\label{fig:hist_Vphi}
	\end{figure*}

	We select the $333\,616$ stellar particles in the SSR with $|Z|\leq 4$~kpc to define the Probability Distribution Function (PDF), $f(V_\phi)$, of the azimuthal velocity by normalizing to 1 the integral of the total sample. We adopt a model based on a Triple Normal Mixture distribution (TNM), namely	
	\begin{equation}
	f(V_\phi) = \sum_{i=1}^{3} w_i N\Big(V_\phi|\langle V_{\phi,i}\rangle, \sigma_{V_{\phi,i}}\Big),
	\label{eq:TNM}
	\end{equation}
	where $0 \leq w_i \leq 1$ is the mixture weight of the $i$-th component and $N$ is a Normal distribution with mean $\langle V_{\rm \phi,i} \rangle$ and standard deviation $\sigma_{V_{\phi,i}}$. The results of the MCMC analysis are listed in Table~\ref{tab:TNM} and visualised in Fig.~\ref{fig:hist_Vphi} (left panel), where the PDF model (red line) appears in good agreement with the observed velocity distribution (blue dots). Further information on the posterior probability of the fitted parameters are reported in Appendix~\ref{app:A2}.
	
	We also tested alternative models with a different number of components. The results for a double Gaussian distribution such as halo+disc or disc+disc are not statistically significant and do not provide good curve fitting of the data. A four-component model, aimed to represent additional disc populations, produced non-significant improvements to the model. 
	
	We note that the weights of the two disc components listed in Table~\ref{tab:TNM} correspond to the total mass of the \textit{young disc} ($\rm{Age} \leq 4$~Gyr) and of the \textit{old disc/thick disc} ($4 \leq \rm{Age} [\rm{Gyr}] \leq 10$), respectively (cfr. Table~\ref{tab:mono-age_hz} considering the different vertical cut applied). The comparison with the mono-age distributions shown in Fig.~\ref{fig:hist_Vphi} (right panel) confirms that these two disc components represent the superposition of the stellar generations formed in the age intervals 0-4 Gyr and 4-10 Gyr. Note that the \textit{old thin disc} and the \textit{thick disc} populations are difficult to distinguish because of their very similar velocity distributions. 
	Moreover, the difference between the mean rotation velocity of the AqC4 discs is
    \begin{equation}
        \Delta V_\phi = 284.4 - 257.7 \simeq 27~\rm{km}~\rm{s^{-1}}.
    \end{equation}
    This value is smaller than in the MW, where $\Delta V_{\rm \phi, MW} = 197.2 - 159.2 = 38~\rm{km}~\rm{s}^{-1}$ \citep{han20}. Actually, this difference depends on the fact that in AqC4 the two main disc components represent the young and the old/thick discs instead of the whole thin disc and the thick disc, as in the MW. Moreover, the high rotation velocity of the AqC4 disc is consistent with its more compact structure with respect to the MW disc (Sect.~\ref{sec:hr}).
	
	Finally, the halo component shows a small prograde rotation as observed in the inner halo of the MW, while its velocity dispersion $\sigma_{V_\phi} \simeq 160$~km~s$^{-1}$ is about $70-100$ per cent higher than in the Solar Neighbourhood \citep[][and references therein]{refiorentin15, bg16}.
	
	In summary, here we robustly support what is discussed in Sect.~\ref{sec:spadist}: in the SSR, we can distinguish at least two kinematic disc components which rotate faster than what is measured in the Solar Neighbourhood (as discussed globally in Sect.~\ref{sec:RC}), and with a smaller discrepancy between the mean values. These results extend the spatial characterisation of AqC4 stellar disc beyond the determination of the scale lengths and scale heights, allowing us to properly investigate the complete phase-space of mono-age populations, as usually done in Galactic surveys.
	
	\begin{table}
		\caption{TNM model parameters: $w_i$ is the mixture weight of the $i$-th component (i.e. young disc, old disc/thick disc, halo), while $\langle V_\phi \rangle$ and $\sigma_{V_\phi}$ are the corresponding mean and the standard deviation of the Normal distribution. Note that $0 \leq w_i \leq 1$ and $\sum_i w_i = 1$.}
		\label{tab:TNM}
		\begin{center}
			\begin{tabular}{clccc}
				\hline
				$i$ & Component & $w$ & $\langle V_\phi \rangle$  & $\sigma_{V_\phi}$ \\
				  & & & [km~$\rm{s}^{-1}$] & [km~$\rm{s}^{-1}$] \\
				\hline
				1. & young disc & 0.389 $\pm$ 0.006 & 284.4 $\pm$ 0.2 & 20.7 $\pm$ 0.2
				\\[+0.5em]
				\multirow{2}{*}{2.} & old disc/ & \multirow{2}{*}{0.467 $\pm$ 0.006} & \multirow{2}{*}{257.7 $\pm$ 0.3} & \multirow{2}{*}{39.9 $\pm$ 0.1} \\ 
				 & thick disc & & & \\[+0.5em]
				3. & halo & 0.145 $\pm$ 0.001 & 35.6 $\pm$ 1.1 & 159.6 $\pm$ 0.6 \\ 
				\hline
			\end{tabular}
		\end{center}
	\end{table}
	
	
\subsection{Kinematics of mono-age disc populations}
\label{sec:mono-agekin}
	
\begin{figure*}
	    \centering
		\includegraphics[width=\linewidth]{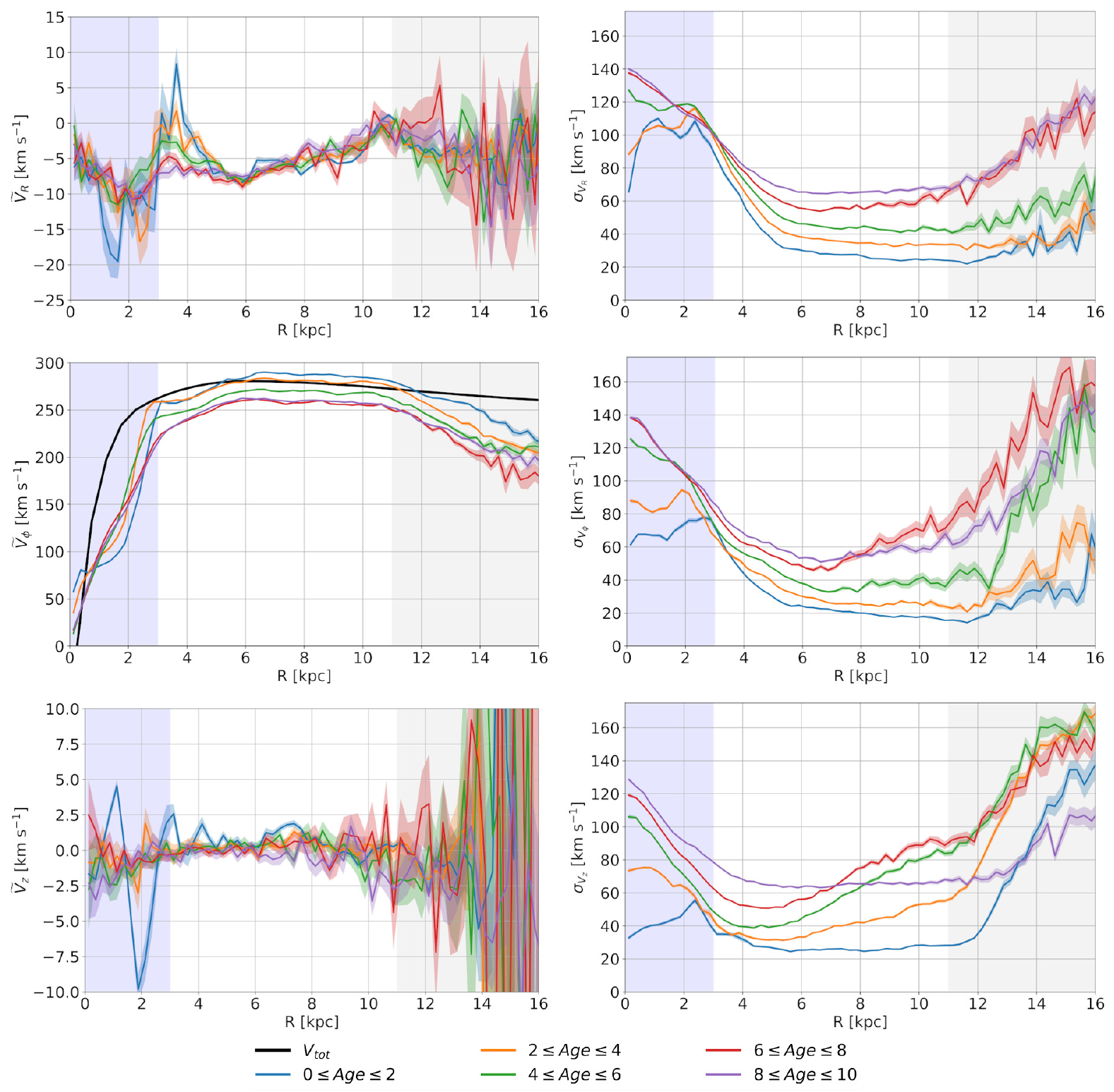}
		\caption{AqC4 median velocity components $\widetilde{V}_R$, $\widetilde{V}_\phi$, $\widetilde{V}_Z$ (left, from top to bottom), and dispersions $\sigma_{V_R}$, $\sigma_{V_\phi}$, $\sigma_{V_Z}$ (right, from top to bottom), as a function of $R$ for different mono-age stellar populations within $|Z| \leq 1$~kpc (same colour code as in Table~\ref{tab:mono-age_hz}). Blue shadow (for $R \leq 3$~kpc) defines central parts of AqC4, while gray shadow (for $R \geq 11$~kpc) stands for halo region. For $3 \leq R [\rm{kpc}] \leq 11$ the disc dominates. Error bars are derived via bootstrapping with 100 re-samples. In panel \emph{left-middle}	we show also the total mass RC (black solid line) for a better comparison.} 
		\label{fig:velocity}
	\end{figure*}

	Fig.~\ref{fig:velocity} shows the three median velocity components $\Big(\widetilde{V}_R$, $\widetilde{V}_\phi$, $\widetilde{V}_Z\Big)$ and the dispersions ($\sigma_{V_R}$, $\sigma_{V_\phi}$, $\sigma_{V_Z}$) as a function of $R$, computed in radial annular bins of $\Delta R = 0.25$~kpc, for the mono-age stellar disc populations younger than 10~Gyr and with $|Z| \leq 1$~kpc. Error bars are derived via bootstrapping with 100 re-samples.
	
	As in Sect.~\ref{sec:RC}, we can identify three main regions based on the different kinematic properties of mono-age populations: (i) for $R \leq 3$~kpc (blue area) we have the bulge/bar region characterized by high velocity dispersion and peculiar velocity patterns; (ii) for $3 \leq R~[\rm{kpc}] \leq 11$ (white area) the region dominated by the disc, where we have the lowest dispersions and the RC flat regime; (iii) finally, for $R \geq 11$~kpc (grey area) the halo region where the velocity dispersions increase and, conversely, the rotation velocity decreases.
	
	In the disc region, the median radial velocity turns out to be negative, i.e. $\widetilde{V}_R = -10 \div 0$~km~s$^{-1}$, for almost all mono-age populations (see Fig.~\ref{fig:velocity}, top-left panel), meaning that the system is out of equilibrium in contrast to what observed in the MW. Examining the last $\sim 0.4$~Gyr dynamic history of AqC4, we found that this global inward motion of the stellar particles characterizes the disc at the present time only, since $\widetilde{V}_R \sim 0$~km~s$^{-1}$ in the previous snapshots of the simulation. Unfortunately, the nature of this kinematical signature cannot be easily investigated because it occurred only in the last snapshot of the simulation. The dynamical perturbation of the AqC4 disc may have been produced by the last satellite detected at redshift $z = 0.014$ (see next section for details), whose high velocity and retrograde orbit shows a perigalactic passage in the outer regions of the disc. Further investigations are necessary to confirm this hypothesis. Despite such zero-point offset in AqC4, we observe an `U-shape' with a minimum at about the Sun position fairly similar to the MW, although the Gaia data samples only a portion of the Galactic disc \citep[Fig.~12 in ][]{k18}. 
	
	The rotation velocity $\widetilde{V}_\phi$ is slower for the older stellar generations, as expected from asymmetric drift, as well as to the secular processes (e.g. spiral arms perturbations and bar resonances) and to merging events. The presence of significant dynamical disc perturbations may also explain why the young stellar particles with $6< R \rm{[kpc]}< 11$ and $|Z| \leq 1$~kpc rotate faster than $V_c$ (middle-left panel). These stars belong to the most angular-momentum sustained population and may be accelerated by the momentum of accreted gas.
	
	As expected by the azimuthal averaging, the radial gradient of the vertical velocity is almost flat, with a median value $\widetilde{V}_Z \simeq 0$~km~s$^{-1}$ and with larger fluctuations for younger populations (Fig.~\ref{fig:velocity}, bottom-left panel). Instead, the apparent radial increase of the mean vertical velocity, $V_Z$, found in the outer disc of the MW by \citet[][ Fig.~3]{poggio18} represents a local signature of the Galactic warp, due to the limited volume sampled by Gaia DR2 and to the peculiar Sun position close to the line of nodes.
	
	In the disc region, the two velocity dispersion components $\sigma_{V_R}$ and $\sigma_{V_\phi}$ decrease from $R \simeq 3$~kpc to $\sim 6$~kpc and then become almost constant until $R \simeq 11$--12~kpc. The younger populations show stronger gradients $\partial\sigma_{V}/\partial R$ and cooler isothermal curves than the older populations. These results are consistent with the monotonic radial decrease of the velocity dispersions observed in our Galaxy, since the colder longer-scale length \textit{young disc} is dominant in the outer disc.  
	
	The behaviour of $\sigma_{V_Z}$ in AqC4 is different from the other two components and shows a positive radial gradient for mono-age populations between 2 and 8~Gyr old (i.e. the intermediate \textit{old disc} population), while the youngest and oldest stellar particles are almost isothermal. We argue that this is a cosmological signature of a heating process due to mergers that occurred in the last 7~Gyr of the simulation as discussed below in Sect.~\ref{sec:SFH}.
	
	Even though our investigation focuses on the disc region, we remark the peculiar velocity patterns within $R \leq 4$~kpc due to the presence of a central bar, which is easily visible in both the stellar and gas distributions (see left panels of Fig.~\ref{fig:AqC4_dens}).
	
	In summary, the stellar disc of AqC4 is still evolving and shows the kinematic signatures of several past and recent dynamical perturbations, such as $(a)$ the global inward $V_R$ systematic motion, $(b)$ the faster rotation $V_\phi > V_c$ of the youngest stars, and $(c)$ the peculiar velocity dispersions shown by mono-age populations. Moreover, the gas accretion in the external region of the disc shown in Fig.~\ref{fig:AqC4_dens} highlights that the galaxy is still out-of-equilibrium, as discussed in \citet{MV20}. All these properties are consistent with the recent studies based on Gaia DR2 that show how the accretion history is the key to understanding the origin of the ancient thick disc and inner halo \citep[e.g.][and references therein]{s13, helmi18}, as well as to disentangle the signatures of the  ``dynamically young and perturbed MW disk'' \citep[][]{antoja18}. 
	
\section{Galaxy formation and evolution}
\label{sec:galevol}
	
	Star Formation History (SFH) does enclose fundamental information about the origin and evolution of disc-like galaxies and of the MW, as well. In order to understand the phase-space properties of mono-age populations discussed in Sects.~\ref{sec:hzSSR}, \ref{sec:TNM} and \ref{sec:mono-agekin}, here we investigate the star formation and the accretion history of AqC4. Similarly to \citet{bignone19} and \citet{grand20}, we contrast the total Star Formation Rate (SFR) with the accretion history of our simulated galaxy. In addition, we consider the local SFR, as inferred by the stellar particles within the SSR, that we compare with the observational results recently derived by \citet{m19}.
	
	
\subsection{Star Formation History}
\label{sec:SFH}
	
	First, we analyse the SFH of the stellar populations in AqC4 by investigating the SFR, which results $\sim~2.65~\rm{M}_\odot \rm{yr}^{-1}$ for the whole galaxy at redshift $z=0$. This value is at least 1.5 times higher than what is measured for the MW by \citet[][]{rw10} and \citet[][]{ln15}.
	
\begin{figure*}
		\centering
		\includegraphics[width=\linewidth]{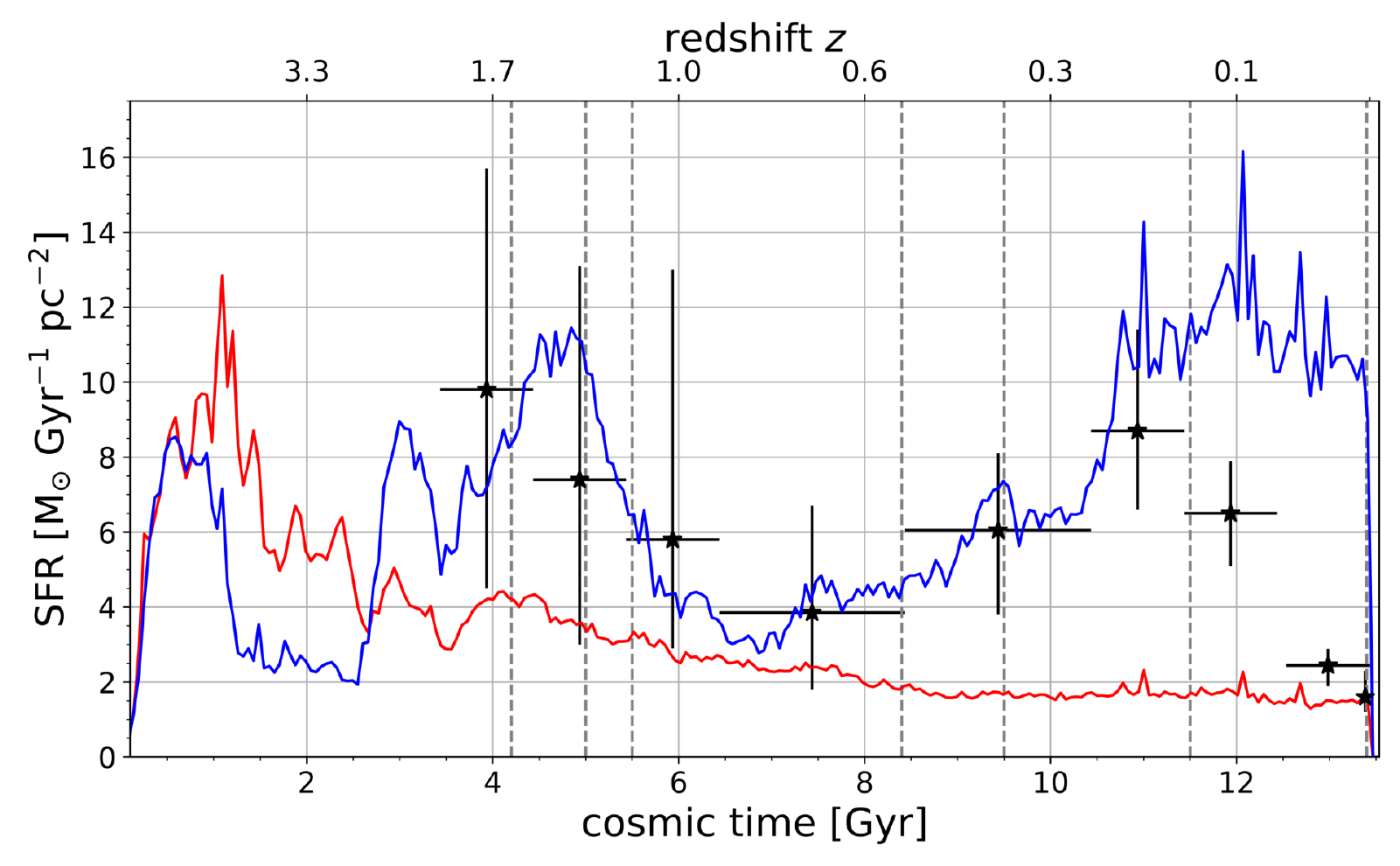}
		\caption{SFR per unit of surface as a function of cosmic time for $r \leq 24.0$~kpc (red line) and in the SSR (blue line). Black symbols refer to data from \citet{m19}. The vertical error bars indicate the 0.16 and 0.84 quantiles of their the posterior estimates and the horizontal error bars indicate the size of the age bin. Vertical dashed lines indicate the redshift/cosmic time of the accreted satellites listed in Table~\ref{tab:mergers}.}
		\label{fig:SFR}
	\end{figure*}

	Fig.~\ref{fig:SFR} shows the evolution of the SFR per unit of surface, for the star particles inside a spherical volume of $R_{\rm gal}$ (red line) and in the SSR with $|Z| \leq 3$~kpc (blue line). Black symbols with corresponding error bars represent the recent stellar production in the Solar Neighbourhood as estimated by \citet{m19} from Gaia DR2.
	
	The apparent peak of the whole SFR at $z > 3.3$ represents the primordial phase of the galaxy formation that builds up the central bulge \citep[as in][]{murante15}, while the smooth decrease after redshift $z\sim 2.5$ (i.e. during the last 10~Gyr) describes the disc formation. 
	
	The SFR of AqC4 shows a secondary peak at redshift $z \sim 1.6$ that corresponds to the thick disc formation discussed in Sects.~\ref{sec:TNM}~--~\ref{sec:mono-agekin}. At this epoch, we estimate the total SFR~$\approx 7$~--~8~M$_\odot$~yr$^{-1}$, which appears more similar to value of $\sim 6.5$~M$_\odot$~yr$^{-1}$ found in the EAGLE simulation studied by \citet{bignone19}, than to the much stronger starburst up to $\sim 25$~M$_\odot$~yr$^{-1}$ resulting from the AURIGA simulation analysed by \citet{grand20}.
	
	Meanwhile, the subset of stellar particles within the SSR highlights an irregular SFH, quite different from the global SFR, that evidences the complex evolution of the galactic disc. The peak at cosmic times $T < 2$~Gyr clearly indicates that a fraction of the stars formed during the primordial starburst have moved from the inner galaxy to the SSR. Then, after a quenching phase, a double-peak event (at redshift $z \sim 2.1$ and $z \sim 1.4$) evidences the formation of the ancient thick disc. We point out the consistency between these bursts in the star formation at $T \sim 4$ -- 6~Gyr and the relative weights of the mono-age populations reported in Table~\ref{tab:mono-age_hz}.
	
	Because of the inside-out formation of the disc, for $T > 7$~Gyr we notice an increasing SFR in the SSR, as opposed to the decreasing SFR of the entire AqC4. Finally, we remark the very high SFR of 10~--~$12~\rm{M_\odot}~\rm{Gyr}^{-1}~\rm{pc}^{-2}$ for $T > 11$~Gyr, that formed a massive young disc, as already discussed in Sect.~\ref{sec:hzSSR}. Despite the large uncertainties, the SFR observed by \citet{m19} in the Solar Neighbourhood results in very good agreement with our estimates for AqC4 in the SSR.
	
	The large time-scale (almost 3.5~Gyr) and the large amount of mass that is involved suggest that this event is produced by an external factor. Indeed, lower panels of Fig.~\ref{fig:AqC4_dens} show a large reservoir of gas both in the disc and falling from the halo. We argue that the low-redshift merging events shown in the next Section may have contributed to the recent gas accretion.
	
	Although more significant in AqC4, this scenario is consistent with the results published by \citet{m19} and supports the hypothesis that the increasing SFR in the Solar Neighbourhood of the MW may be due to the recent merging event claimed by \citet{lian20}.
	
\subsection{Accretion history}
\label{sec:merger}
	
	In order to understand the features found in the SFH of AqC4, we select seven snapshots with redshift $z = 1.637$, $1.314$, $1.154$, $0.550$, $0.401$, $0.178$ and $0.014$, corresponding to the time intervals of the main starbursts observed in Fig.~\ref{fig:SFR}.
	
\begin{figure*}
	\centering
		\includegraphics[width=\linewidth]{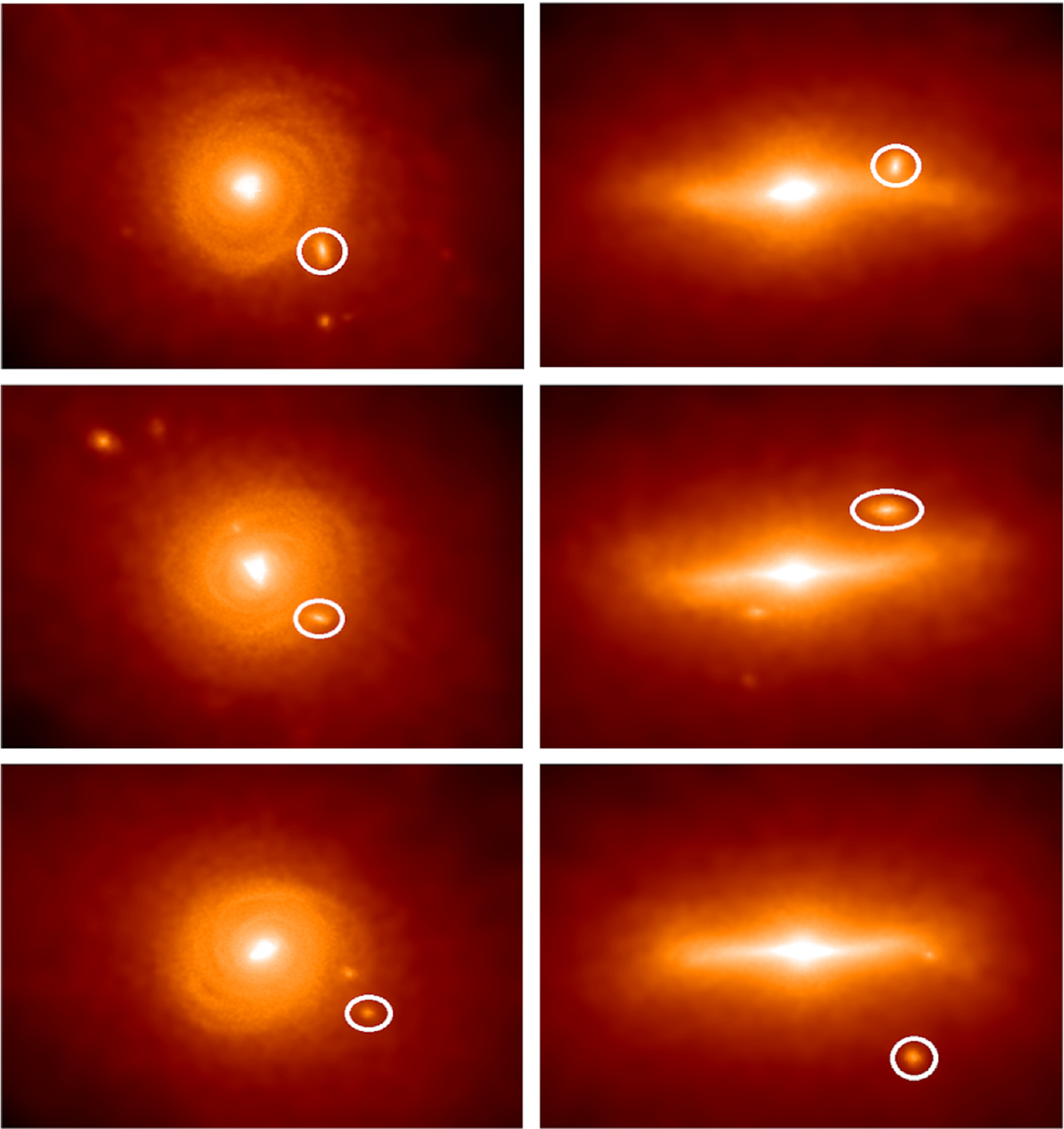}
		\caption{Stellar density maps for the AqC4 simulation at redshift $z = 1.637$, $1.314$ and $1.154$, from top to bottom. Left-hand panels show face-on projections, right-hand panels show edge-on projections. White circles highlight the selected merging satellites.}
		\label{fig:snap_maps}
	\end{figure*}

\begin{figure*}
	\centering
		\includegraphics[]{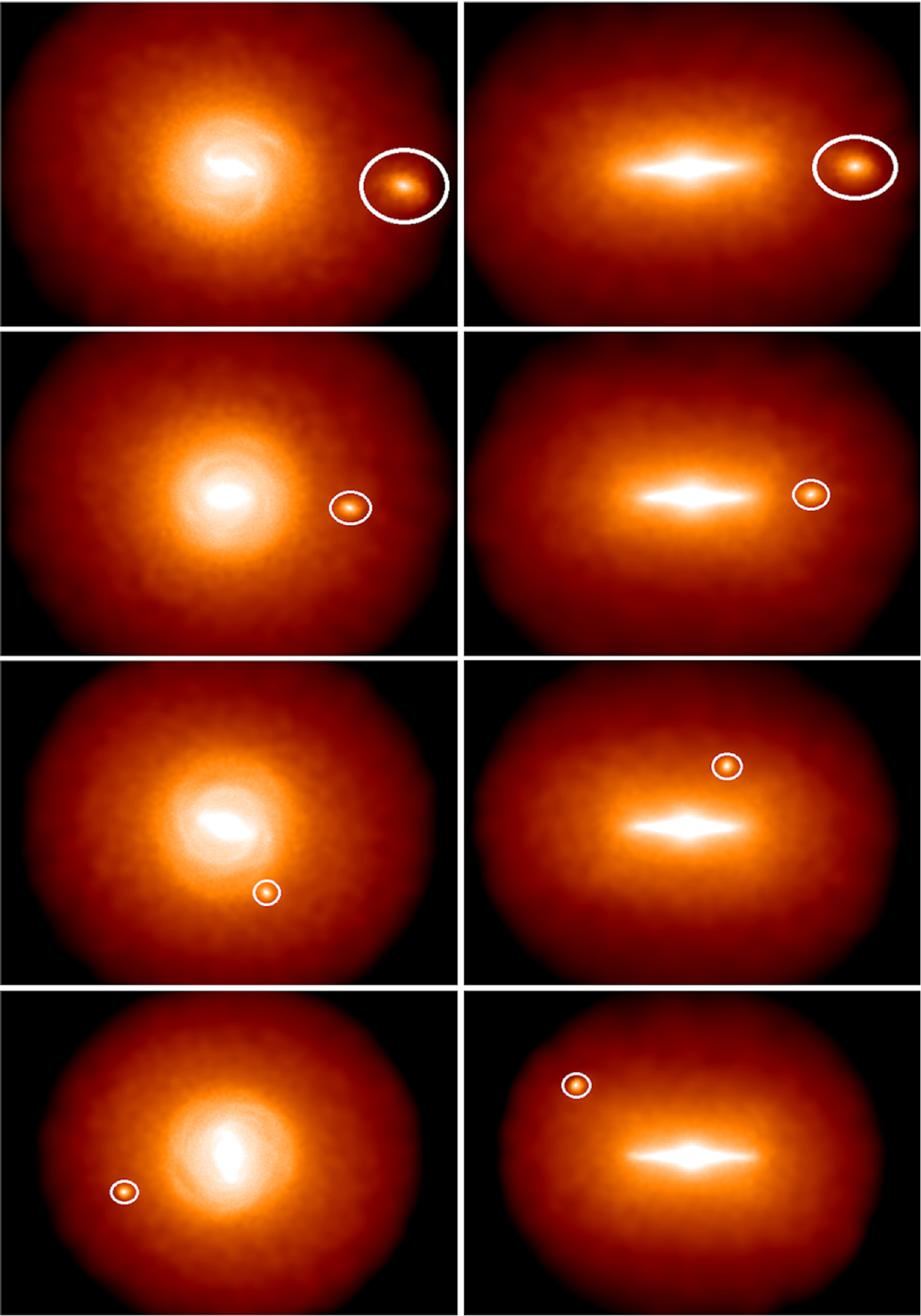}
		\caption{As Fig.~\ref{fig:snap_maps} but at redshift $z = 0.550$, $0.401$, $0.178$ and $0.014$.}
		\label{fig:snap_maps2}
	\end{figure*}

	For illustration purpose only, in Figs.~\ref{fig:snap_maps}~-~\ref{fig:snap_maps2} we show the face-on and edge-on stellar density distributions at the selected redshifts. The presence of one or more merging satellites is noticeable in all the panels.
	
	For each snapshot, we estimate the total stellar mass of the main satellite, marked with a white circle, within a spherical radius, $r_{\rm sat}$. A similar procedure is performed to estimate the mass of the main disc-like galaxy, but considering a cylindrical volume defined by the radial disc extension at that epoch, $R_{\rm disc}$, within $|Z| < 2$~kpc. As reported in Table~\ref{tab:mergers}, the stellar mass ratio $\Delta = M_*^{\rm sat}/M_*^{\rm disc}$ varies from $\sim 0.5$ to 5.5 per cent.

	We focus our attention on the satellite at redshift $z = 1.637$, whose stellar mass of $1.2 \cdot 10^9~\rm{M_\odot}$ is pretty similar to that estimated by \citet{mac19} and \citet{f19} for the GSE progenitor. In fact, these authors claim a mass of a few $10^9~\rm{M_\odot}$, assuming a $10^{10}~\rm{M_\odot}$ stellar mass for the thick disc present at the time. We obtain a stellar mass ratio $\Delta \sim 5.5$ per cent which is quite consistent with the stellar mass ratio of $\sim 6$ per cent estimated by \citet{helmi18} and confirmed by \citet{gal19} for GSE. This event is the main merger that can be associated with the starburst between 4 and 6~Gyr in cosmic time (see Fig.~\ref{fig:SFR}).
	
	After this large merger, we find a few accretion events due to satellites with masses of $M_*^{\rm sat} \lesssim 10^8~\rm{M_\odot}$ corresponding to a mass ratio $\Delta \leq 2$ per cent. These minor accretions produced the secondary fluctuations of the SFR in the SSR after redshift $z \sim 1.5$. These results are consistent with the standard hierarchical formation scenario of the MW \citep[][and references therein]{helmi20}.
	
	We estimate a mass of $M_*^{\rm sat} \simeq~8.8~\cdot~10^8~\rm{M_\odot}$ for the accreted satellite detected at $z=0.550$, and a mass of $M_*^{\rm sat} \simeq~5.1~\cdot~10^8~\rm{M_\odot}$ for the one detected at redshift $z = 0.401$. These low-redshift significant mergers may have produced an heating of the \textit{old disc} population that explains the large scale height of the stellar particles with age 4-8 Gyr shown in Fig.~\ref{fig:flaring}. The presence of merging events during the intermediate phase of the disc formation is consistent with the late-accretion model proposed by \citet{lian20}, who suggest that a recent merger event occurred in the MW at $8.2$~Gyr in cosmic time (i.e. redshift $z \sim 0.6$). In particular, these authors estimate a mass of $M_*<10^9~\rm{M_\odot}$ for the gas-rich dwarf galaxy involved.
	
	Finally, we detect two more satellites at redshift $z = 0.178$ and $0.014$ , with M$_*^{\rm sat} \simeq 4.0~\cdot~10^8$~M$_\odot$ and $2.6~\cdot~10^8$~M$_\odot$, respectively, which appear associated to the increasing SFR in the SSR of AqC4 during the last 2~Gyr.
	
	In summary, the SFH and the accretion history of AqC4 are consistent with its spatial and kinematic properties described in Sects.~\ref{sec:spadist}~--~\ref{sec:kinprop} and, in particular, they clarify the formation of the peculiar features observed in the \textit{young disc} and \textit{old disc} populations.
	
	Moreover, the overall formation and evolution of AqC4 and the MW appear fairly similar. Indeed, our findings are consistent with both the GSE scenario for the origin of the Galactic \textit{thick disc} \citep{brook04, brook12, s13, helmi18, gal19}, and with the recent accretion event proposed to explain the increasing SFR in the Solar Neighbourhood \citep{m19}. We argue that the higher SFR of our simulated galaxy with respect to the MW depends on both the gas contribution from the late accreted satellites and the infall of gas previously expelled by the strong starbursts and supernova explosions that occurred at high redshift.
	
	\begin{table}
		\caption{Estimates of the total stellar mass of selected satellite galaxies and the main galactic disc, with corresponding mass ratios $\Delta = M_*^{\rm sat}/M_*^{\rm disc}$ at different redshift $z$. The spherical radius, $r_{\rm sat}$, is centred on satellite galaxies, while the radial extension of the disc, $R_{\rm disc}$, is in the reference frame of the main galaxy.}
		\label{tab:mergers}
		\begin{center}
		\begin{tabular}{c|cc|cc|c}
			\hline
			$z$ & $r_{\rm sat}$ & $M_*^{\rm sat}$ & $R_{\rm disc}$ & $M_*^{\rm disc}$ & $\Delta$ \\
			 & [kpc] & $[10^8 \rm{M_\odot}]$ & [kpc] & $[10^{10} \rm{M_\odot}]$ & $[\%]$ \\ 
			\hline
			1.637 & 3.0 & 12.0 & 5.0 & 2.2 & 5.5 \\  
			1.314 & 2.5 & 4.7 & 5.0 & 2.7 & 1.7 \\  
			1.154 & 4.0 & 1.3 & 5.0 & 2.9 & 0.4 \\
			0.550 & 6.0 & 8.8 & 8.0 & 4.5 & 2.1 \\
			0.401 & 3.5 & 5.1 & 8.5 & 4.8 & 1.1 \\
			0.178 & 3.5 & 4.0 & 9.0 & 5.3 & 0.8 \\
			0.014 & 3.5 & 2.6 & 10.0 & 5.7 & 0.4 \\
			\hline
		\end{tabular}%
		\end{center}
	\end{table}
	
\section{Summary and Conclusions}
\label{sec:concl}
	
	In this work, we analysed the spatial and kinematic properties of stellar particles of the MW-like galaxy AqC4. Our approach consisted in considering such simulation as a real survey of the stellar contents of our Galaxy. Therefore we implemented methods of investigation usually applied to real stellar catalogues. Our aim was to identify cosmological signatures enclosed in the mono-age stellar populations, defined as coeval particles sub-samples of 2~Gyr age bins, and compare their phase-space properties with the MW. We focused on the Simulated Solar Ring (SSR), an annular region within $6 < R {\rm [kpc]} < 7$, corresponding approximately to the Solar Neighbourhood, given the $\sim 20$ per cent smaller radial scale length of the thin disc of AqC4 with respect to the MW ($h_{\rm R} \simeq 2.1$~kpc wrt. 2.6~kpc), and where we were able to compare the spatial-kinematical distributions of the stellar populations to the SFH and accretion history of this simulated galaxy.
	
	The high-resolution cosmological simulation we used is based on the Aquila-C suite presented by \citet{spr08} and the MUPPI algorithm to model the sub-resolution baryonic physics \citep[see ][ for more details]{murante10, murante15}. It is an unconstrained simulation with initial conditions chosen to reproduce a disc-like galaxy. Therefore, it should be interpreted as a cosmological product with mass and phase-space properties similar to the MW. The Plummer-equivalent softening length for the computation of the gravitational force is $\epsilon_{\rm Pl} = 163 h^{-1}$~pc.
	
	In the SSR, we confirmed the presence of at least two disc components with vertical exponential distribution (see Sect.~\ref{sec:hzSSR}). We estimated a scale height $h_1 \sim 0.3$~kpc that is in good agreement with the MW thin disc, and a scale height $h_2 \sim 1.33$~kpc that is $\sim 50$ per cent larger with respect to the MW thick disc. 
	
	The inspection of the vertical distributions of the mono-age stellar particles in the SSR clarified that the thin disc of AqC4 is mainly formed by a \textit{young disc} population with $h_{\rm Z} \sim 250$ -- 500~pc and age $\leq 4$~Gyr (Sects.~\ref{sec:hzSSR}~-~\ref{sec:flaring}). In addition, we identified an \textit{old disc} (age 4-8~Gyr) having intermediate thickness ($h_{\rm Z} \sim 500$ -- 1000~pc), and a prominent \textit{thick disc} population with age 8-10~Gyr and $h_{\rm Z} \sim 1200$~pc (Fig.~\ref{fig:flaring}).
	
	We evidenced that the vertical scale heights and weights of the mono-age populations are strictly correlated with the SFH of AqC4. In fact, the large scale heights of the stellar particles in the SSR with age 4-8~Gyr cannot be explained by the secular disc evolution, and we argued that these values derive from the disc heating produced by the low-redshift mergers shown in Fig.~\ref{fig:SFR}.
	
	The kinematic analysis of the stellar particles in the SSR (Sect.~\ref{sec:TNM}) revealed that the azimuthal velocity distributions correspond to two main kinematic disc components (i.e. the young disc and the old/thick disc), plus a slightly prograde inner halo. We pointed out that these discs rotate faster than the MW, and with a smaller velocity difference, $\langle V_{\phi,1} \rangle \simeq 284.4$~km~s$^{-1}$ and $\langle V_{\phi,2} \rangle \simeq 257.7$~km~s$^{-1}$. These results are consistent with the more compact structure of AqC4 with respect to the MW.
	
	The median $\widetilde{V}_R$ and $\widetilde{V}_\phi$ components of mono-age stellar particles and their relative dispersions in Fig.~\ref{fig:velocity} revealed that the stellar disc of AqC4 is out of equilibrium and shows the dynamical signatures of the perturbations due to both recent mergers, cosmic impact and gas accretion. Indeed, in the disc region $3 \leq R [\rm{kpc}] \leq 11$, $\widetilde{V}_R$ evidences a systematic inward motion for all the mono-age populations in contrast to what measured in the MW. We suppose that the very recent impact of the high-speed, counter-rotating satellite found at redshift $z = 0.014$ (see lower panels of Fig.~\ref{fig:snap_maps2}) may have produced this dynamical perturbation. On the other hand, younger stars show an higher median $\widetilde{V}_\phi$ than older ones as expected from inside-out formation and secular processes occurred in the disc, but also rotate faster than the circular velocity, accelerated by accreated gas.
	
	These kinematic features are consistent with the recent studies based on Gaia DR2 data that show how the accretion history is the key to understand the origin of the ancient MW thick disc and inner halo \citep[e.g.][]{brook04, brook12, s13, helmi18, gal19}, as well as to disentangle the signatures of the  ``dynamically young and perturbed MW disk'' \citep[][]{antoja18}.
	
	We suggest that the prominent \textit{thick disc} population was generated by the major accreted satellite detected at redshift $z = 1.6$. Such merging event is associated to the starburst at redshift $1 < z < 2$ shown in Fig.~\ref{fig:SFR}, which attains a total SFR of $\sim 7$~--~8~M$_\odot$~yr$^{-1}$. This scenario is consistent with the results published by \citet{bignone19} and \citet{grand20} who analysed independent cosmological simulations of MW-like galaxies selected from the EAGLE and Auriga projects, respectively. Although such simulations show SFH's quite different from AqC4, all these studies evidence a significant increase in the SFR triggered by the galaxy merger and associated to the thick disk formation.
	
	We also compared the age distribution of the disc stellar particles selected within the SSR with the SFH in the Solar Neighborhood. At redshift $1 < z < 2$, we detected an apparent signature of the starburst above discussed, which matches quite well the high SFR ($\sim 10$~M$_\odot$~Gyr$^{-1}$~pc$^{-2}$) recently derived at the same epoch by \citet{m19}. The SFR in the SSR is fairly similar to that derived for the MW by \citet{m19} till $z \simeq 0.2$. The comparison with the accretion history of AqC4 supports the hypothesis that the starburst that occurred 2~--~4~Gyr ago in the Solar Neighbourhood may be due to a late merging event, as claimed for the MW at $z \sim 0.6$ by \citet{lian20}.
    
    The higher SFR of our simulated galaxy in the SSR with respect to the MW at low redshift may depend on a greater gas contribution from the last accreted satellites, as well as from the delayed infall of the gas outflows triggered by the strong starbursts at high redshift.
	
	Given this, our analysis sheds light on the complex scenario of the origin and evolution of the Galactic disc: it supports the global top-down, inside-out disc formation model and implies an overlapping of several stellar generations closely correlated with the accretion history of the Galaxy. 
	
    The negative median $\widetilde{V}_R$ trends represents an interesting starting point for future studies that will have to include chemo-dynamical analysis in order to improve and detail our knowledge of the MW thin and thick discs origin and evolution, as well as investigations on the \emph{in-situ/ex-situ} star formation contributions and stellar back-time tracking in order to select halo streams and identify common progenitors to study the Galactic halo. In this respect, the use of constrained simulation \citep[e.g.][]{carlesi16} can be very promising.
	
	There is no doubt that current and future Gaia data releases and the important synergies with spectroscopic ground surveys such as APOGEE \citep{apogee} and GALAH \citep{galah} and detailed comparison with simulations will bring tremendous and fundamental contributions to the studies of Galactic Archaeology.
	
	\section*{Acknowledgements}

    We would like to thank the anonymous referee and the Assistant Editor for their thoughtful comments and suggestions that helped us improve our manuscript. We thank also Eloisa Poggio and Ronald Drimmel for the useful discussions.
    
    Simulation was carried out using ULISSE at SISSA and Marconi at CINECA, Italy (project IsB16\_DSKAGN, PI:G. Murante). The post-processing has been performed using the PICO HPC cluster at CINECA through our expression of interest. We thank Volker Springel for making the GADGET3 code available to us. This research made use of python libraries scipy \citep{scipy}, corner \citep{corner}, and PyMC3 \citep{pymc3}.
	
	We are indebted to the Italian Space Agency (ASI) for their continuing support through contract 2018-24-HH.0 to the National Institute for Astrophysics (INAF). MV is supported by the Excellence Cluster ORIGINS, which is funded by the Deutsche Forschungsgemeinschaft (DFG, German Research Foundation) under Germany's Excellence Strategy - EXC-2094 - 390783311.
	
	\section*{Data availability}
	
	The data underlying this article will be shared on reasonable request to the corresponding author.








\bibliographystyle{mnras}
\bibliography{bibliography}



%
	

\appendix


\section{Posteriors PDF}
\label{app:A}

\subsection{Vertical distribution of stellar disc}
\label{app:A1}

\begin{figure}
	    \centering
	    \includegraphics[width=\linewidth]{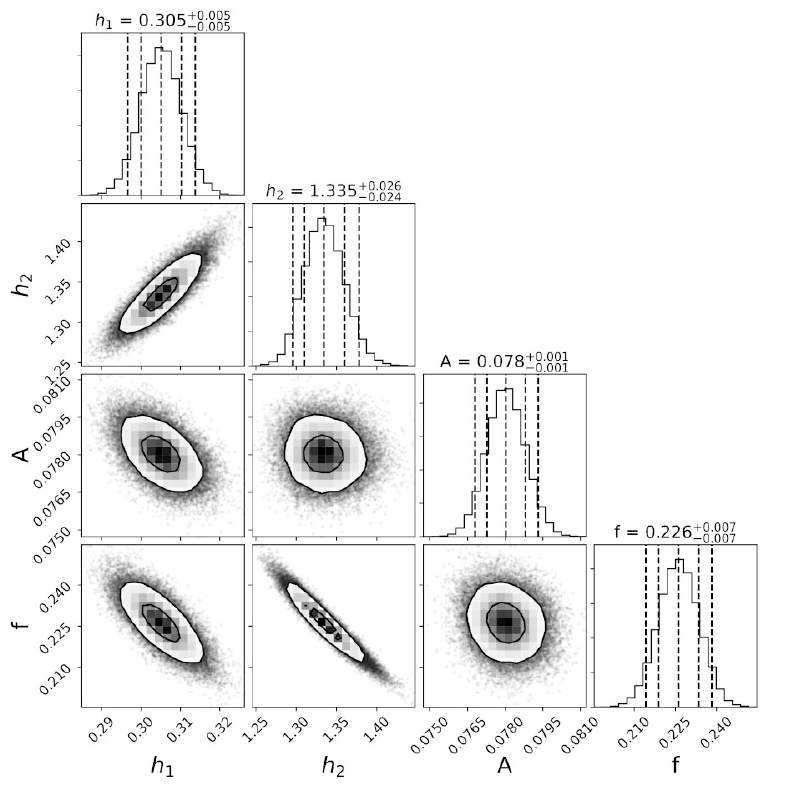}
	    \caption{Posterior distributions of the MCMC analysis for the vertical stellar particles distribution in the SSR. The one-dimensional (histogram) posterior distributions for each parameter are shown on the diagonal, while the other panels represent the two-dimensional (contours) correlations.}
	    \label{fig:post_hz}
    \end{figure}

    Fig.~\ref{fig:post_hz} shows the posterior distributions of the parameters according to Eq.~(\ref{eq:rhoZ}) for the vertical distribution of stellar particles. Dashed lines in each histogram refer to the 10$^{\rm{th}}$, 16$^{\rm{th}}$, 50$^{\rm{th}}$ (i.e. median), 84$^{\rm{th}}$ and 90$^{\rm{th}}$ percentiles of the relative distribution, while numbers on top indicate the medians and the 1~$\sigma$~CIs. Thick black contours indicate the 1 and 2~$\sigma$~CI of the two-dimensional correlations of the posteriors.
    
    As for the radial scale length, the uncertainties take into account the Poissonian statistic within each bin and are smaller than the size of single data points in the plot. Consequently, the CIs are very narrow and we obtain well-peaked posteriors on the parameters estimates.

    The analysis highlights that $h_1$ and $h_2$ are positively correlated, and they are both negatively correlated with the local density normalization parameter $f$ indicating the intrinsic physical overlapping of the two disc components. The closest correlation is between $h_2$ and $f$.

\subsection{TNM model}
\label{app:A2}

\begin{figure*}
	    \centering
	    \includegraphics[width=\textwidth]{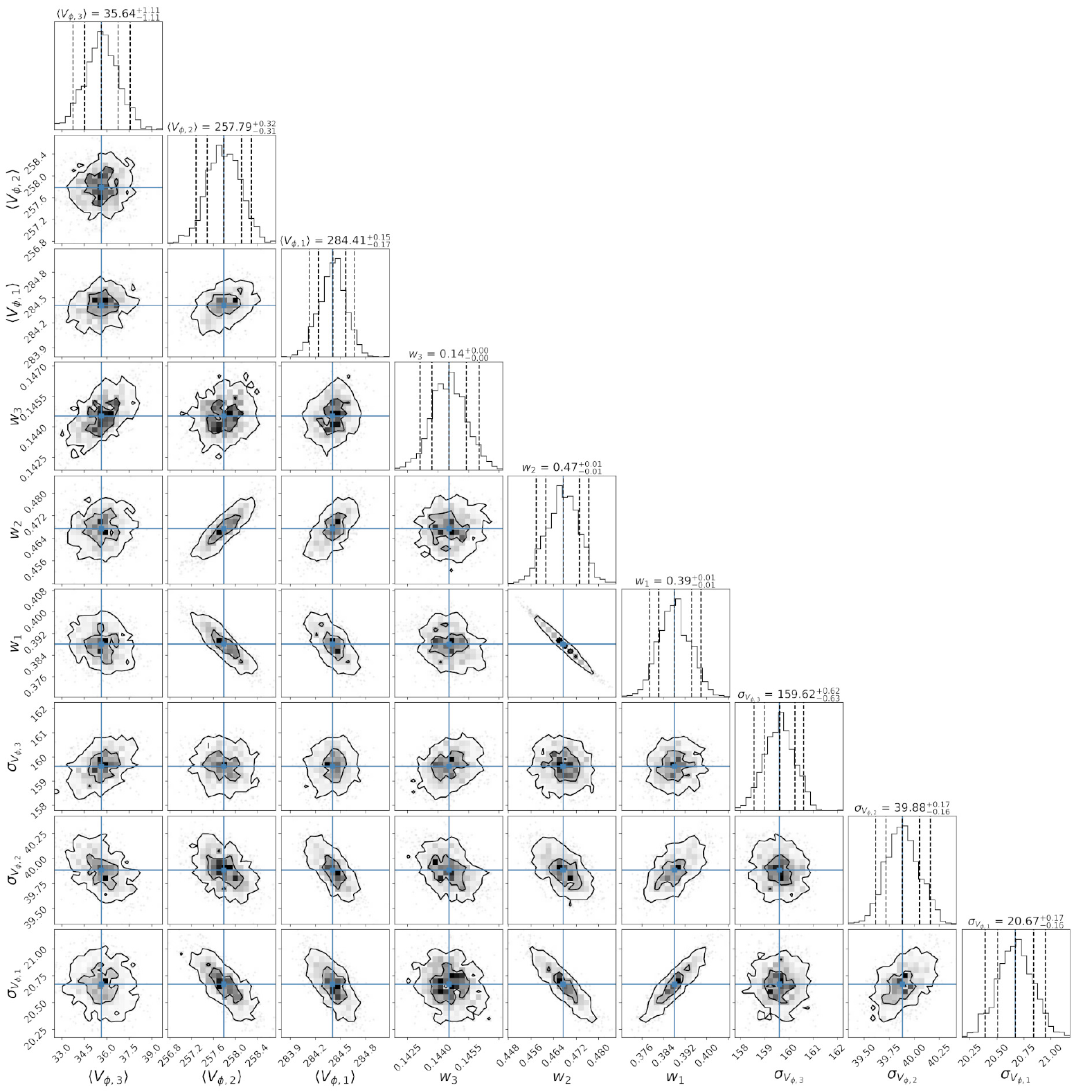}
	    \caption{As in Fig.~\ref{fig:post_hz} for TNM model parameters according to Eq.~\ref{eq:TNM}. The blue square shows the mean value of each posterior distribution.}
	    \label{fig:corner_Vphi}
    \end{figure*}

    The posterior distributions of the parameters for TNM model are shown in Fig.~\ref{fig:corner_Vphi}. The means of the posteriors are indicated with a blue square, while dashed black lines and numbers on top of each histogram have the same meaning as in Appendix~\ref{app:A1}. The posteriors are well approximated by normal distribution, as the means and the medians are similar and the CI are pretty symmetric.

    The closest correlations are between the arrays of parameters that describe the two discs components, i.e. the components with $i=1,2$. This is again a signature of the intrinsic overlap of the stellar generations that compose the two populations and is similar to what reported in Appendix~\ref{app:A1}. The last kinematic component with $i=3$ represents the residual halo contribution and its parameters show almost no correlations with the other quantities.

\bsp	
\label{lastpage}

\end{document}